\documentclass[showpacs,preprintnumbers,amsmath,amssymb,prd,twocolumn,
superscriptaddress]{revtex4}
\usepackage{graphicx}% Include figure files
\usepackage{dcolumn}% Align table columns on decimal point
\usepackage{bm}% bold math
\def\comment#1{}

\begin{document}

\title{Phase Structure of $d=2+1$ Compact Lattice Gauge Theories and 
the Transition from Mott Insulator to Fractionalized Insulator.}
\author{J. Smiseth}
\email{jo.smiseth@phys.ntnu.no}
\affiliation{Department of Physics, Norwegian University of
Science and Technology, N-7491 Trondheim, Norway}
\author{E. Sm{\o}rgrav}
\email{eivind.smorgrav@phys.ntnu.no}
\affiliation{Department of Physics, Norwegian University of
Science and Technology, N-7491 Trondheim, Norway}
\author{F. S. Nogueira}
\email{nogueira@physik.fu-berlin.de}
\affiliation{Institut f{\"u}r Theoretische Physik, 
Freie Universit{\"a}t Berlin,
D-14195 Berlin, Germany}
\author{J. Hove}
\email{joakim.hove@phys.ntnu.no}
\affiliation{Department of Physics, Norwegian University of
Science and Technology, N-7491 Trondheim, Norway}
\author{A. Sudb{\o}}
\email{asle.sudbo@phys.ntnu.no}
\affiliation{Department of Physics, Norwegian University of
Science and Technology, N-7491 Trondheim, Norway}

\date{Received \today}

\begin{abstract}
Large-scale Monte Carlo simulations are employed to study phase transitions
in the three-dimensional compact abelian Higgs model in adjoint representations 
of the matter field, labelled by an integer $q$, for $q=2,3,4,5$. We also study 
various limiting cases of the model, such as the $Z_q$ lattice gauge theory, dual 
to the $3DZ_q$ spin model, and the $3DXY$ spin model which 
is dual to the $Z_q$ lattice gauge theory in the limit $q \to
\infty$. 
In addition, for benchmark purposes, we study the 
square lattice $8$-vertex model,
which is exactly solvable and features non-universal critical exponents.
We have computed the first, second, and third moments of the action to
locate 
the phase-transition of the  compact abelian Higgs model in the
parameter 
space $(\beta,\kappa)$,
where $\beta$ is the coupling constant of the matter term, and $\kappa$ 
is the coupling constant of the gauge term. We have found that for $q=3$, 
the three-dimensional compact abelian Higgs model has a phase-transition 
line $\beta_{\rm{c}}(\kappa)$ which is first order for $\kappa$ below a finite 
{\it tricritical} value  $\kappa_{\rm{tri}}$, and second order above. 
The $\beta=\infty$ first order phase 
transition persists for finite $\beta$ and joins the second order phase transition 
at a tricritical point 
$(\beta_{\rm{tri}}, \kappa_{\rm{tri}}) = (1.23 \pm 0.03, 1.73 \pm 0.03)$.
For all other integer $q \geq 2$ we have considered, the entire phase-transition 
line $\beta_c(\kappa)$ is critical. We have used finite-size scaling of the 
second and third moments of the action to extract critical exponents $\alpha$ 
and $\nu$ without invoking hyperscaling, for the $XY$ model, the $Z_2$ spin and 
lattice gauge models, as well as the compact abelian Higgs model for $q=2$ and 
$q=3$. In all cases, we have found that for practical system sizes, the third 
moment gives scaling of superior quality compared to the second moment. We have 
also computed the exponent ratio for the $q=2$ compact $U(1)$ Higgs model 
along the critical line, finding {\it a continuously varying ratio} 
$ (1+\alpha)/\nu$, as well as continuously varying $\alpha$ and $\nu$ as 
$\kappa$ is increased from $0.76$ to $\infty$, with the Ising universality 
class $(1+\alpha)/\nu = 1.763$ as a limiting case for 
$\beta \to \infty, \kappa \to 0.761$, and the $XY$ universality class 
$(1+\alpha)/\nu = 1.467$ as a limiting case for $\beta \to 0.454,\kappa \to \infty$. 
However, the critical line exhibits a remarkable resilience of $Z_2$ criticality 
as $\beta$ is reduced along the critical line. Thus, the three-dimensional compact 
abelian Higgs model for $q=2$ appears to represent a {\it fixed-line} theory 
defining a new universality class. We relate these results to a recent microscopic 
description of zero-temperature quantum phase transitions within
insulating phases of 
strongly correlated systems in two spatial dimensions, proposing the above to be 
the universality class of the zero-temperature {\it quantum phase 
transition} from a Mott-Hubbard insulator to a charge-fractionalized insulator in 
two spatial dimensions, which thus is that of the $3D$ Ising model for a considerable 
range of parameters.  
\end{abstract}


\maketitle
\section{Introduction}
Lattice gauge theories in $2+1$ dimensions with compact gauge-fields 
coupled to matter fields, have recently come under close scrutiny as 
effective theories of strongly correlated fermion systems in two spatial
dimensions at zero temperature. Phase transitions in such three dimensional
models then correspond to quantum phase transitions in a 
system at zero temperature in two spatial dimensions. A central issue is whether 
such systems of strongly correlated fermions can suffer quantum phase 
transitions from Fermi-liquid metallic states to states where the quasiparticle 
concept has broken down and given way to singular Fermi liquids  \cite{Review} 
or electron-splintered states \cite{Senthil,Kivelson}. Such quantum phase 
transitions may be related to phase transitions such as 
confinement-deconfinement transitions in $2+1$ dimensional compact gauge 
theories. This fact has resulted in focused attention  on effective gauge 
theories of matter fields representing charge doped into Mott-Hubbard 
insulators, coupled to fluctuating gauge fields representing strong 
constraints on the dynamics of the fermions on the underlying lattice 
on which the models are defined 
\cite{IntroRef,Matsui,Laughlin,DHLee,Nagaosa}. Since these are 
lattice models, the corresponding gauge fields are necessarily 
compact. Compact $U(1)$ gauge fields in $2+1$ dimensions support stable 
topological defects in the form of monopole configurations, and it has been 
suggested that the unbinding of such monopoles, a confinement-deconfinement 
transition, may be relevant for such phenomena as spin-charge separation 
in strongly correlated systems \cite{Mudry,Matsui,Nagaosa}. Confinement 
here refers to the confinement of test charges in the problem, 
not of topological defects of the gauge field (which are space-time instantons, 
and will hereafter be referred to as ''monopole'' configurations). Alternative
formulations in terms of a lattice Ising gauge theory \cite{Wegner} coupled to 
matter fields, have also been put forth \cite{RS,Wen,Senthil}. This is largely 
motivated, it would appear, by the fact that it is highly controversial whether 
a $2+1$-dimensional $U(1)$ gauge theory with matter fields {\it in the fundamental 
representation} will undergo a confinement-deconfinement transition. In the absence 
of matter fields, compact $U(1)$ gauge theories are known to be permanently confined 
in $d=2+1$, while the pure $Z_2$ gauge theory is known to have a second order phase 
transition in the inverted $3D$ Ising universality class. It is sometimes stated, 
without much justification, that the presence of matter fields in the fundamental 
representation will not alter the picture that emerges in their absence, both 
for the $U(1)$ case and the $Z_2$ case in $2+1$ dimensions.  

One major problem that arises in this context, is that the Wilson loop non-local 
gauge-invariant order parameter, which has proven itself to be very useful in the 
absence of dynamical matter-fields to distinguish confined from deconfined phases, 
is rendered  useless by the presence of them. In particular, when the 
symmetry group of the matter field is contained in the symmetry group of 
the gauge field, one can demonstrate rigorously under otherwise quite general
conditions, that the Wilson loop is 
bounded from below by a perimeter law, not an area law, despite the fact 
that models with this property definitely has phase transitions from confined 
to deconfined phases \cite{Nussinov}. Hence, the Wilson loop and the related 
Polyakov loop, are no longer useful order parameters for the problem. Physically, 
this is due to the fact that in the presence of a dynamically fluctuating 
matter-field coupled to the gauge field, particle-hole excitations are generated 
from the vacuum and will effectively screen the interaction between two static 
test-charges introduced into the system, which the Wilson loop is a measure of. 
Hence, a perimeter law is always obtained.

Recently, two of us have shown that a confinement-deconfinement transition may take 
place in three-dimensional compact $U(1)$ gauge theory coupled to matter fields, when 
the matter field exhibits {\it critical} fluctuations \cite{KNS,KNS1}. Such matter-field 
fluctuations endow the gauge-field with an anomalous scaling dimension $\eta_A$ with  
a value  \cite{Herbut,Hove}, 
\begin{eqnarray}  
\eta_A = 4 - d,
\label{eta_A}
\end{eqnarray}
where $d$ is the dimensionality of the system. This value for $\eta_A$ is protected by 
gauge-invariance. 
In $d=3$, this transforms the gauge-field propagator in a striking manner such as to allow 
a confinement-deconfinement transition to take place via a {\it three-dimensional} 
Kosterlitz-Thouless like phase-transition of unbinding of point-like monopole configurations 
of the gauge-field \cite{KNS,KNS1}.  The treatment of Refs. \onlinecite{KNS,KNS1} closely parallels 
that of Ref. \onlinecite{Polyakov} and is based on a dual description in the 
continuum. However, it is far from obvious what the corresponding {\it lattice gauge 
theory}, if any, that would yield such results, could be. 

In the absence of a clear-cut order-parameter criterion for distinguishing 
various phases of such matter-coupled gauge fields, it would be advantageous to 
be able to distinguish various phases by direct ''thermodynamic'' measurements 
reliably exhibiting possible non-analytic behavior. In this paper, we will use
one such measurement recently introduced by us \cite{sudbo1}, namely finite-size 
scaling of  the {\it third moment of the action of the lattice model}. This will 
turn out to be a superior quantity to study for this purpose, compared to the second 
moment. It brings out non-analytic thermodynamics and precision values of the specific 
heat critical exponent $\alpha$ and the correlation length exponent $\nu$ through 
finite-size scaling analysis performed on practical system sizes \cite{sudbo1}. 
The third moment has the advantage of not being contaminated by contributions from 
the second moment. The second moment is known to be a notoriously difficult quantity to 
use in particular for extracting specific heat exponents, due to large corrections 
to scaling coming from confluent singularities for practical system sizes.  Moreover, 
the third moment is the simplest quantity to compute which has an extra feature which 
even moments do not have. It has a double-peak structure where the width {\it between} 
the peaks also exhibits scaling. This allows us to extract separaely {\it two} 
exponents, $\alpha$ and $\nu$, from measurements of the third moment alone, without 
having to invoke hyperscaling. 

A lattice  model of particular interest in this context is the abelian $U(1)$ Higgs
model with a compact gauge-field \cite{Polyakov} coupled minimally to a $U(1)$ 
bosonic matter field \cite{Savit,FradShe} with a gauge-charge $q$. It is defined 
by the partition function given by the following functional integral 
%\begin{widetext}
\begin{eqnarray}
Z & = & \int_{-\pi}^\pi\left[\prod_{j=1}^{N} \frac{d \theta_j}{2\pi}
\right]\int_{-\pi}^\pi\left[\prod_{j,\mu} \frac{d A_{j\mu}}{2\pi}\right]  
 \exp \left[\beta  ~ H_\beta +  \kappa ~ H_\kappa \right] \nonumber \\
H_\beta &=& \sum_{j, \mu} (1-\cos(\Theta_{j \mu}))  \nonumber \\
H_\kappa &=& \sum_{{\rm P},\mu} (1-\cos({\cal A}_{j \mu})),
\label{Model1}
\end{eqnarray}
%\end{widetext}
where $N$ is the number of lattice sites, and we have defined 
\begin{eqnarray}
\Theta_{j \mu} & = & \Delta_{\mu} \theta_j-q A_{j\mu}  \nonumber \\ 
{\cal A}_{j \mu} & = & \varepsilon_{\mu\nu\lambda} \Delta_{\nu} A_{j\lambda}.
\end{eqnarray}
Here, $\varepsilon_{\mu \nu \lambda}$ is the completely antisymmetric tensor.
Moreover, $\sum_{j,\mu}$ denotes a sum over sites of the lattice, while 
$\sum_{\rm{P}, \mu}$ denotes a sum over the plaquettes of the lattice.  We will use 
the variables $(x=1/[\kappa +1],y=1/[\beta+1])$ when discussing the possible phases 
of this model \cite{Savit}. In Eq. (\ref{Model1}), $\theta$ is the phase of a 
scalar matter field with unit norm representing holons, $\Delta_{\mu}$ is a 
forward lattice difference operator in direction $\mu$, while $A_{j \mu}$ is the 
fluctuating gauge field enforcing the onsite constraints reflecting the strong 
correlations in the problem. We are neglecting amplitude fluctuations of the 
matter fields, working in the ``London"-limit. 

Let us summarize what is known about this model. When $q=0$, the matter field  
decouples from the gauge field. It is well known that the model then has  one 
critical point in the universality class of the $3DXY$ model, and in the Villain 
approximation (which is most often used when dualizing the model \cite{Savit}), 
the critical value 
$y_c$ is given by $y_c \approx 3/4$ \cite{Villain,Note}. On the other hand, the pure gauge 
theory is permanently confined for all values of $\kappa$ \cite{Polyakov}. Consider 
next $q=1$. Then, Eq. (\ref{Model1}) is trivial on the line $x=1,0<y<1$  with no 
phase transition  for any value of $y$. On the line $0<x<1,y=1$ the matter field 
is absent and the theory is permanently confined \cite{Polyakov}. For a further 
enumeration on  rigourous results both on the non-compact and compact version of 
this model, see also \cite{Borgs}.

For arbitrary $(\beta,\kappa)$, the case $d=3,~ q=1$ is controversial. It is 
however clear that no ordinary second order phase transition with a local 
order parameter exists for the model in this case. When matter-fields are coupled 
to a compact gauge-field in a continuum theory, the permanent confinement of 
the pure gauge theory is destroyed and a confinement-deconfinement transition 
may take place via a Kosterlitz-Thouless like unbinding of monopole configurations 
\cite{KNS} in {\it three dimensions}. This is due to the appearance of an anomalous 
scaling dimension of the gauge-field induced by critical matter-field 
fluctuations \cite{Herbut,Hove}. The role of the anomalous dimension has
also been 
studied recently at finite temperature, in pure compact QED in $d=3$
with 
no matter fields present \cite{Chernodub1}. In both
Refs. \onlinecite{KNS} 
and \onlinecite{Chernodub1}, the 
anomalous scaling dimension plays a crucial hole. However, in Ref. 
\onlinecite{Chernodub1} the deconfinement transition occurs due to finite 
temperature, and there are no matter fields.  

The role of an anomalous scaling dimensions in driving a recombination of monopole 
defects of the gauge field into dipole configurations, corresponding
to a 
confinement-deconfinement transition, has been studied numerically for the case $q=1$ in the 
presence of matter-fields \cite{Chernodub2}. The authors of Ref. \onlinecite{Chernodub2} 
in this case reach conclusions in agreement with Ref.  \onlinecite{KNS}. Thus, even 
putting the issue of whether this signals a phase-transition or not aside, it is 
clear that also matter fields in the fundamental representation {\it do} matter in 
this problem. It is also interesting to note in this context, that some time ago a 
rather remarkable paper \cite{Amit}, a Kosterlitz-Thouless transition 
was claimed in a three-dimensional theory of integer point charges interacting via a 
logarithmic potential. There however, the origin of the logarithmic interaction was 
due to higher order anisotropic gradient terms, essentially an input to the theory, and 
not a {\it result} of  an anomalous scaling dimension appearing due to criticality in the 
mass-sector of a compact $U(1)$ gauge theory. This however,
contrasts with the case considered in Ref. \onlinecite{KNS}. The anisotropy effectively leads 
to a dimensional reduction and a resulting standard Kosterlitz-Thouless phase transition 
in two dimensions. 

For $q>1$ the model in Eq. (\ref{Model1}) exhibits a behavior that completely 
sets it apart from the case $q=1$. The reason is that for $q >1$, the matter 
field is no longer in the fundamental representation. Thus, in contrast to 
the no-compact case of the standard Maxwell term of the gauge sector, the 
gauge charge cannot be simply scaled away. The case $q=2$ is particularly 
interesting in the context of electron fractionalization \cite{Motrunic}. 
This is a case that we will consider in detail in this paper. This theory 
arises as a special limit of a bosonic model exhibiting fractionalized phases 
considered recently \cite{Motrunic}. The phase diagram for $q=2$ was briefly 
discussed long ago by Fradkin and Shenker \cite{FradShe}. For $d=3$ the 
phase diagram is divided into two phases, a confined and a 
deconfined-Higgs phase. There is no Coulomb phase in three 
dimensions. The model becomes a $Z_2$ gauge theory on the line $y=0$. 
This limit suffices to bring out the fundamental difference between the 
$q=1$ and $q=2$ cases, since on the line $y=0$, the $q=1$ case
is trivial. We will discuss this in more detail below. The vortex 
content of the $q=2$ case is different: $Z_2$ vortices, or 
{\it visons}, may arise in the deconfined-Higgs phase. Due to the 
visons the flux is quantized in units of $2\pi$ instead of 
$\pi$. This means that the excitations in the deconfined-Higgs 
phase have charge $e/2$. In the context of the bosonic 
model in Ref. \onlinecite{Motrunic}, this phase corresponds to a 
{\it fractionalized insulator}. The confined phase, on the 
other hand, features excitations with charge $e$ and should 
correspond to a conventional Mott insulator \cite{Motrunic}. 
Thus, the model (\ref{Model1}) can be thought as describing a 
particular case of {\it insulator-fractionalized insulator} 
transition. The full bosonic model considered in Ref. 
\onlinecite{Motrunic} has a more rich phase structure depending 
on the values of the parameters. For example, a superfluid 
phase is also possible. We will not consider such a situation 
here, but only a special limit of the bosonic model of 
Ref.\onlinecite{Motrunic}. 

The purpose of this paper is to make a detailed numerical 
study of the phase structure of (\ref{Model1}). An important 
point concerns the universality class of the phase transition. 
This will be the main topic discussed in this paper. 
We have performed a large scale Monte Carlo 
study which gives a very complete picture of the universality 
class of the transition. The present study complements and 
goes far beyond our previous large scale Monte Carlo study of 
the model \cite{sudbo1}. In Ref. \onlinecite{sudbo1}, we have shown that the 
Ising ($Z_2$) universality class dominates over a significant portion 
of the critical line of the theory. 
Moreover, we have shown that there is a region of parameters where 
the critical exponents are continuously varying. This was interpreted 
as evidence for the existence of a {\it line of fixed points} in the 
renormalization group (RG) flow diagram. To the portions of 
the critical line corresponding respectively to $Z_2$ and 
$XY$ critical behavior, are associated Ising and $XY$ fixed points. 
Besides these two fixed points there exists a fixed line corresponding to 
a {\it critical phase}. As discussed recently \cite{Kivelson}, electron 
fractionalization in $1+1$ dimensions is associated to a quantum critical 
phase. Ordinarily, in more than one spatial dimension quantum phase transitions 
are associated with a critical point. Thus, we would expect that electron 
fractionalization would occur at a critical point. {\it Our analysis clearly 
shows that a {\it critical phase} exists for the model (\ref{Model1}) at} 
$q=2$. This becomes particularly clear on the portion of the critical line 
corresponding to large values of $\kappa$, where we find that the critical
exponents vary smoothly with coupling constants,
approaching the $3DXY$ value only slowly.  The three-dimensional 
KT-like scenario \cite{KNS,KNS1} for the $q=1$ model is also an example of  
critical phase occuring in higher dimensions. However, due to the 
vortex content of the model at $q=1$, it is not entirely obvious
that the corresponding deconfinement transition is really associated with 
electron fractionalization \cite{Nayak}. 

In this paper the cases $q=3,4,$ and $5$ will also be considered. 
The case $q=3$ is particularly interesting. While for $q=2,4,$ and $5$ 
the {\it entire} line separating the two phases are critical, this is 
not true when $q=3$. In this case there is a value of 
$\kappa$ below which the transition is first-order, being second-order 
otherwise. This point where the transition changes from second order 
to first order, is clearly a tricritical point. We emphasize that among 
the situations analysed in this paper, only the case $q=3$ exhibits a 
tricritical point. When $q=3$, charge is fractionalized in such a 
way that excitations carry charge $e/3$. This situation is 
reminiscent of the $\nu=1/3$ state in the fractional quantum 
Hall effect \cite{DasSarma}. In general the charge for arbitrary 
$q$ will fractionalize as $e/q$. This gives an elementary flux 
quantum $\phi_0=2\pi q/e$, contrasting with the more familiar 
situation of flux quantization in a superconductor. Indeed, if we 
take the example of the $q=2$ theory, we see that the flux quantum 
is doubled, while in a superconductor it is halved because the 
Cooper pair has charge $2e$. Other interesting possibilities 
of vortex/charge fractionalization were considered recently 
in certain Josephson junction arrays \cite{Doucot}. 

In all these examples, it is readily seen that the type of flux quantization 
involved depends on specifying whether charge is fractionalized 
or not. Therefore, in order to detect charge fractionalization, 
experimental measurements of flux as the one proposed in Ref. 
\onlinecite{SFtop-order} in the context of underdoped cuprates 
are important. The experiment proposed in Ref. \onlinecite{SFtop-order} 
has now been performed  \cite{Bonn}, but has failed 
in  detecting electron fractionalization. The failure 
of this experiment does not, however, entirely rule out charge 
fractionalization in the cuprates for the following deep reason. 
In the zero doping insulating regime, the system can  
effectively be described by the antiferromagnetic quantum Heisenberg 
model. This model is known to have a {\it local} $SU(2)$ gauge 
symmetry \cite{Zou}. Upon doping, this symmetry is broken 
down to $U(1)$. Since this $U(1)$ is a subgroup of $SU(2)$, 
it is compact and carries the ``charge'' label $q$ corresponding 
to the different representations. Within the $q=2$ represention, 
the double broken symmetry pattern $SU(2)\to U(1)\to Z_2$ 
is possible. In such a situation, there would be no stable 
vison and any experimental attempt based on a vison trapping 
experiment would fail. However, this does not mean that charge is 
not fractionalized. The $Z_2$ residual subgroup will imply 
the existence of magnetic monopoles which, by duality, accounts for 
the existence of fractional charges $e/2$. 

In the perspective of such a possible scenario, alternative
experimental proposals should be envisioned and the identification 
of the universality class of possible phase transitions the $q=2$ 
case is important in this context.            

The plan of the paper is the following. In Section II, we discuss 
the limits of the model, pointing out the differences with respect 
to the $q=1$ case. In Section III, we explain the finite-size 
scaling analysis of the third moment of the free energy used 
in this paper. In Section IV we discuss the details of our 
large-scale Monte Carlo simulations. As a check on the quality 
of the simulations, we have performed several benchmark simulations
which we present in parallel with our new results on the model
defined by Eq. (\ref{Model1}). Section V concludes the paper.      
    
\section{Limits of the compact gauge theory}

\subsection{$\beta \to \infty$, $\kappa$ finite}
In terms of the formulation of the model in Eq. (\ref{Model1}), this limit 
leads to the constraint $\Delta_{\mu} \theta_j - q A_{j\mu} = 2 \pi l_{i,\mu}$ 
where $l_{i,\mu}$ is integer valued. Substituting this into the gauge-field 
term, we find   
%\begin{widetext}
\begin{equation}
Z = \prod_{j=1}^{N} \sum_{l_{j,\mu}=-\infty}^{\infty}
 \exp \left[\kappa \sum_{{\rm P},\mu} (1-
 \cos (\frac{2 \pi}{q} {\cal L}_{j,\mu} ))\right], 
\label{Model2}
\end{equation}
%\end{widetext}
%where $\sum_{\rm{P},\mu}$ denotes a sum over plaquettes of the lattice.
where ${\cal L}_{j,\mu} = \varepsilon_{\mu\nu\lambda} \Delta_{\nu} l_{j,\lambda}$ is also 
integer valued.  For $q=1$ the model is therefore trivial in this particular limit, in 
accordance with the discussion in the previous section. For $q=2$, the model is equivalent 
to the lattice $Z_2$ gauge theory \cite{Wegner}, and the critical point of the model in 
this limit is thus in  the inverted Ising universality class (in analogy with the inverted 
$XY$  universality class of the dualized $3DXY$ model \cite{Various}). Therefore $\alpha$ and 
$\nu$ are those of the global 3D Ising model. For integer $q \geq 3$, the  critical exponents 
$\alpha$ and $\nu$ we will find in the limit $\beta \to \infty$, $\kappa$ fixed, will be 
those of the $Z_q$ spin model, since as we shall see below, the $Z_q$ gauge theory is 
dual to the $Z_q$ spin model in $d=3$. 

For arbitrary integer-valued gauge-charge $q$ (i.e. labelling of matter-field
representation where $q=1$ means fundamental representation, 
while $q \geq 2$ means 
higher representations), we may write the action in Eq. (\ref{Model1}) in the  
Villain approximation, replacing the cosine-terms by periodic quadratic parts, 
after which the model may be written in terms of its topological defects as 
follows \cite{Savit,Nagaosa}
\begin{widetext}
\begin{equation}
Z  = Z_0\sum_{\{Q_j\}} \sum_{\{ J_{j\nu} \}} 
\delta_{\Delta_{\nu} J_{j\nu},q Q_j}
\exp \left[-4 \pi^2 \beta \sum_{j,k} ~ \biggl( J_{j\nu} 
~J_{k\nu} + \frac{q^2}{m^2} Q_jQ_k \biggr)D(j-k,m^2)\right],
\label{Loopgas0}
\end{equation}
\end{widetext}
where $D(j-k,m^2)=(-\Delta_{\lambda}^2 + m^2)^{-1} ~\delta_{jk}$ and $m^2=q^2 \beta/\kappa$. 
Here $Z_0$ is the partition function for massive spin waves \cite{Savit}, and is an analytic function 
of coupling constants which will be omitted from now on. Note the appearance of the constraint 
\begin{eqnarray}
\Delta_{\nu} J_{j\nu} = q Q_j;~~ q \in {\mathbb{N}}
\label{Constraint}
\end{eqnarray}
in the summation, which will be important in what follows.
Here, $Q_j \in\mathbb{Z}$ 
is the monopole charge on (dual) lattice site number $j$, while $J_{j \nu}$ are topological 
currents representing segments of open-ended strings terminating on monopoles, or closed 
loops. These are the only stable topological objects of the theory \cite{Savit}. For a 
recent treatment of the  interplay between abelian monopole condensation and vortex 
condensation in lattice gauge theories, see \cite{Cea}.
 
In the limit $\beta \to \infty$ at fixed $\kappa$, the partition function 
in Eq. (\ref{Loopgas0}) takes the form
\begin{equation}
Z=  \sum_{\{Q_j\}} \sum_{\{ J_{j\nu} \}}   
\delta_{\Delta_{\nu} J_{j\nu},q Q_j}\exp
\left(-\frac{2 \pi^2 \kappa}{q^2}  \sum_{j} ~  J^2_{j\nu}\right).
\label{Theory}
\end{equation}
This is easily seen to be the loop-gas representation of the global 
$Z_q$ theory in the Villain approximation \cite{Savit1}. Using the 
integral representation of the Kronecker delta and summing over 
$J_{j\nu}$ using the Poisson formula, we obtain up to an overall 
non-singular factor

\begin{widetext}
\begin{equation}
Z=\sum_{\{N_{j\nu}\}}
\int_{-\pi}^\pi\left[\prod_{j=1}^{N} \frac{d \theta_j}{2\pi}
\right]\exp\left[-\frac{q^2}{8\pi^2\kappa}\sum_{j,\nu}
(\Delta_\nu\theta_j-2\pi N_{j\nu})^2+iq\sum_j Q_j\theta_j\right].
\end{equation}
\end{widetext}
By employing the Poisson identity

\begin{equation}
\sum_{Q=-\infty}^\infty e^{iqQ\theta}=2\pi\sum_{l=-\infty}^\infty
\delta(\theta-2\pi l/q),
\end{equation}
we obtain

\begin{equation}
\label{ZqVillain}
Z=\sum_{\{l_j=-q+1\}}^{q-1}\sum_{\{N_{j\nu}\}}
\exp\left[-\frac{q^2}{8\pi^2\kappa}\sum_{j,\nu}
\left(\frac{2\pi}{q}\Delta_\nu l_j-2\pi N_{j\nu}\right)^2\right].
\end{equation}
The above is precisely the Villain form of a $Z_q$ model \cite{Savit1}. 

Since Eqs. (\ref{Theory}) and (\ref{Model2}) are dual (up to a Villain 
approximation), and Eq.  (\ref{Theory}) is a loop-gas representation of the 
global $Z_q$ theory while Eq. (\ref{Model2}) is the $Z_q$ lattice gauge theory,
we conclude that the global and local $Z_q$ theories are dual to each other in 
$d=3$, in agreement with the results of Ref. \onlinecite{Guth}. Hence, again 
we conclude that the model Eq. (\ref{Model1}) in the limit $\beta \to \infty$,  
$\kappa$ fixed, should have critical exponents  $\alpha$ and $\nu$ consistent 
with the $q$-state clock model universality class. Since 
$\lim_{q \to \infty} Z_q = U(1)$, the above fits nicely in with what is known 
for the $U(1)$ case in $d=3$, where the global $3DXY$ model dualizes into a $U(1)$ 
gauge theory \cite{Various,KBook,Tesanovic,Nguyen}. 

From Eq. (\ref{Theory}), it is seen that the cases $q=1$ and $q \neq 1$ 
are fundamentally different. For $q=1$, the summations over $\{Q_i\}$ may 
be performed to produce a unit factor at each of the $N$ dual lattice site, 
thus completely eliminating the constraint \cite{Note}. Hence, we have
$Z=\left(\vartheta_3(0,e^{-2 \pi^2 \kappa}) \right)^N$ where the elliptic 
Jacobi functions are given by 
$\vartheta_3(z,q)=\sum_{n=-\infty}^{\infty} q^{n^2} \exp(2 \pi i n z)$, 
which are analytic functions and hence no phase-transition  occurs at any value 
of $\kappa$ for $q=1$ in this limit. For (integer) $q \geq 2$, a phase transition 
is known to survive \cite{FradShe}. The restoration of a phase-transition for 
integer $q > 1$ is, in this language, crucially dependent on the presence 
of the constraint Eq. (\ref{Constraint}) for $q  \neq 1$. Even when one sums 
over {\it all} possible values of $\{ Q_j \}$, this still represents a real 
constraint on the vortex-configurations of the system, since it cannot be 
eliminated by summation as in the case of $q=1$. This suffices to convert the 
theory to a strongly interacting one capable of sustaining a phase-transition,
in contrast to the effectively (discrete Gaussian) noninteracting case $q=1$.

\subsection{$\beta$ finite, $\kappa \to \infty$}
We now discuss the limit $\beta$ finite, $\kappa \to \infty$.  Note from Eq.
(\ref{Model1}), that when $\kappa \to \infty$, we have 
$\varepsilon_{\mu\nu\lambda} \Delta_{\nu} A_{j\lambda}=2 \pi M_{j,\mu}$,
where $M_{j,\mu}$ is integer valued. This shows that in this limit, the 
gauge-field fluctuations are precisely like transverse phase-fluctuations in 
the $3DXY$-model \cite{Hove1}, and the integer $M_{j,\mu}$ plays the role 
of vorticity. Thus, even when $\kappa \to \infty$, gauge-field fluctuations 
are not completely suppressed, we are not allowed to set $M_{j,\mu}=0$, and 
hence we cannot set $A_{j,\mu}=\Delta_{\mu} \chi_j$ in the first term in 
Eq. (\ref{Model1}). However, we may write the partition function on the form
%\begin{widetext}
\begin{equation}
Z = \int_{-\pi}^\pi\left[\prod_{j=1}^{N} \frac{d \theta_j}{2\pi}
\right]\int_{-\pi}^\pi\left[\prod_{j,\mu} \frac{d A_{j\mu}}{2\pi}\right]  
 \exp \left[\beta H_\Theta \right], 
\label{Model3}
\end{equation}
%\end{widetext}
where the functional integral over the gauge-field is now constrained. Introducing 
the Villain-approximation, using the Poisson summation formula, integrating over 
$\theta_j$, and performing a partial integration, leads to the partition function
%\begin{widetext}
%\begin{equation}
%Z = \prod_{j=1}^{N} \sum_{M_{j,\mu}=-\infty}^{\infty} \sum_{L_{j,\mu}=-\infty}^{\infty}
% \exp \left[- 
% \sum_{{j}, \mu}~\frac{1}{2\beta}
%(\varepsilon_{\mu \nu \lambda} \Delta_{\nu,j} L_{\lambda,j})^2 + 2 \pi i q 
%~ L_{\mu,j} ~ \varepsilon_{\mu \nu \lambda} \Delta_{\nu,j} M_{\lambda,j}  
%\right], 
%\label{Model4}
%\end{equation}
%\end{widetext}
which is equivalent to an ordinary loop gas with completely suppressed
gauge-field fluctuations when $q$ is an integer. The gauge-field fluctuations 
of the compact sector produce closed vortex loops with long-range interactions 
such that the loop-tangle is incompressible, entirely equivalent to ordinary 
matter-field vortex loops. {\it No screening of their test charges is produced 
by these constrained gauge-field fluctuations} and monopole configurations are 
suppressed. Hence, the model in this limit is, after all, equivalent to the 
$3DXY$ model, as it also is in the non-compact case where gauge-field 
fluctuations are entirely suppressed. 

Ideally, to explore the entire phase diagram of the model Eq. (\ref{Model1}), we could 
carry out Monte Carlo (MC) simulations in conjunction with finite-size scaling analysis 
of standard quantities such as the susceptibility or the specific heat. The problem with 
the former quantity is that we have no candidate gauge-invariant order parameter for 
this model, in view of the results of \cite{Nussinov,Note_S,Sedgewick}. 
The problem with the latter is that the second moment of the action is marred 
with huge corrections to scaling and is hence unsuitable to extract critical 
exponents from simulations on practical system sizes. To circumvent this 
difficulty, we will consider various higher order {\it moments} of the action 
appearing in the partition function in order to obtain quantities with good 
scaling for practical system sizes we can handle in Monte Carlo simulations.

%The problem 
%with the latter is that, since we do not know {\it a priori} what the sign of the specific 
%heat exponent of the model of Eq. (\ref{Model1}) is for arbitrary couplings $\beta,\kappa$,
%we are faced with the possibility that $\alpha$ in fact could be negative
%also in  the interior of the phase-diagram defined by the parameters 
%$x=1/(1+\kappa),y=1/(1+\beta)$. Direct finite-size scaling  of
%the specific heat, or more generally, the second moment of the
%Hamiltonian for the case with more than one coupling constant, is 
%problematic when $\alpha < 0$ because subdominant singular contributions
%to the free energy representing corrections to scaling dominate
%the results for practical system sizes. Subtraction of such subdominant 
%contributions to the free energy and the specific heat is difficult in 
%the absence of analytical results. 

\section{Moments of the action}
Given a singular contribution to the free energy $F = -\ln Z$ of a system with 
an inverse temperature coupling $\beta$, \begin{eqnarray}
F_{\rm{sing}}  \sim |t|^{2-\alpha} ~~ {\cal{F}}_{\pm}(h/|t|^\Delta), 
\end{eqnarray}
where $h$ plays the role of a scaling variable,  $\Delta$ is some scaling exponent,
$\lim_{x \to \infty} {\cal{F}}_{\pm}(x) \sim x^{(2-\alpha)/\Delta}$,
$\lim_{x \to 0}{\cal{F}}_{\pm}(x) = A_{\pm }^{\cal F}$,
$A_{\pm }^{\cal F}$ are critical amplitudes,
and $t=(\beta-\beta_c)/\beta$ 
is some deviation from a critical coupling, then the non-analytic 
contribution to the susceptibility of the action is given by 
%\begin{eqnarray}
%C = \frac{\partial^2 F_{\rm{sing}}}{\partial \beta^2}, 
%\end{eqnarray} 
%thus
\begin{eqnarray}
C \sim |t|^{-\alpha} ~~ {\cal{C}}_{\pm}(h/|t|^\Delta), 
\label{2smom}
\end{eqnarray}
where $\lim_{x \to \infty} {\cal{C}}_{\pm}(x) \sim x^{-\alpha/\Delta}$,
and $\lim_{x \to 0} {\cal{C}}_{\pm}(x) = A_{\pm}^{\cal{C}}$.
%For simplicity, we have written down the scaling functions for the 
%cases where a critical line in the $(\beta,\kappa)$ phase diagram 
%has been crossed by varying only one of the coupling constants, 
%so that the field $h$  represents the coupling constant which 
%is not being varied across the transition line. 

At a critical point, this quantity will scale 
as $C \sim L^{\alpha/\nu}$, where the volume of the system is given 
by $L \times L \times L$, provided the system exhibits one diverging 
length scale at the phase-transition. Here, $\nu$ is the correlation 
length critical exponent, $\xi \sim |t|^{-\nu}$ close to a critical point. 
A problem arises if $\alpha < 0$, as in the $3DXY$ model, since one then  
gets an increasing peak in $C$ itself with increasing $L$, which however 
eventually will no longer scale with $L$. Unfortunately, but quite typically,
impractically large system sizes are needed to eventually distinguish 
corrections to scaling from actual scaling in the second moment, particularly 
so when $\alpha <0$. Thus, $C$ exhibits a 
finite cusp non-analyticity which does scale, superposed on a large 
regular background which eventually will not. It would be advantageous to 
be able to subtract out this background, or at the very least bring out 
the leading non-analyticity dominating the scaling more clearly relative 
to confluent singularities, or corrections to scaling. This can 
effectively be achieved by taking one further derivative of the action 
with respect to the coupling constant as follows
\begin{eqnarray}
\frac{\partial^3 F_{\rm{sing}}}{\partial \beta^3} 
\sim |t|^{-(1+\alpha)} ~~ {\cal{G}}_{\pm}(h/|t|^\Delta), 
\label{3smom}
\end{eqnarray}
where the scaling function ${\cal{G}}_{\pm}(x)$ has the following properties
$\lim_{x \to \infty}{\cal{G}}_{\pm}(x) \sim x^{-(1+\alpha)/\Delta}$,
and $\lim_{x \to 0} {\cal{G}}_{\pm}(x) \to A_{\pm}^{\cal{G}}$,
which will scale as $L^{(1+\alpha)/\nu}$ at a critical point. 

More generally, the $n$-th moment
\begin{eqnarray}
\frac{\partial^nF_{\rm{sing}}}{\partial \beta^n} 
\sim |t|^{-(n-2+\alpha)} ~~ {\cal{G}}_{\pm}(h/|t|^\Delta), 
\label{nsmom}
\end{eqnarray}
will scale as $L^{(n-2+\alpha)/\nu}$ at a critical point, and therefore by
computing two moments, say the third $(n=3)$ and the fourth $(n=4)$, it is 
also possible to extract $\alpha$ and $\nu$ separately without utilizing the 
hyperscaling relation $\alpha = 2 - d \nu$. 

{\it In fact, this may be obtained from the third moment $M_3$ alone}, since the
width between the negative and positive peaks scales as $L^{-1/\nu}$. Thus
$M_3$ yields independent measurements of both $(1+\alpha)/\nu$ and $1/\nu$.
The above procedures may serve as {\it checks} on the validity of hyperscaling. 
This is known to be violated  above the upper critical dimension in spin-models 
and systems with long-range interactions due to the presence 
of dangerous irrelevant variables \cite{Cardy}. Long-range interactions are features 
of the {\it dual} models of some of the theories we consider and little is known 
about the presence of dangerous irrelevant operators in some of the models we 
study in this paper. Hence, caution is necessary in extracting individual exponents.
At any rate, the above seems to be a useful way of extracting {\it two} exponents
$\alpha,\nu$ from measurements of {\it one} quantity, namely $M_3$. The procedure
described above will presumably turn out to be useful in a host of other models in 
statistical physics. 

\begin{figure}[htbp]
\centerline{\scalebox{0.50}{\rotatebox{0.0}{\includegraphics{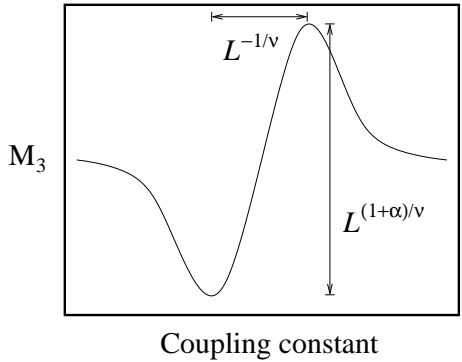}}}}
\caption{\label{Fig1} 
Generic third moment of action, $M_3$, showing how finite-size scaling is used to 
extract $\alpha$ and $\nu$.}
\end{figure}

Consider next a model with two coupling constants $x_1$ and $x_2$ defined by its action 
\begin{eqnarray}
S = x_1 ~ H_1 + x_2 ~ H_2. 
\label{two_coup_action}
\end{eqnarray}
In this case, the scaling of $\frac{\partial^2 F}{\partial x_i \partial x_j}$ could conceivably 
depend on the {\it direction} in parameter space in which the critical line is crossed. To take this complication into account, we consider contributions to the free energy to second order in devitations in the coupling constants from their critical values, by expanding to second order around the critical point  
\begin{eqnarray}
F &=& -\ln{Z}  \nonumber\\
&=& F(\{x_c\}) +  \left.\frac{1}{2}
\sum_{i,j}\delta x_i \delta x_j \frac{\delta^2 F}{\delta x_i \delta x_j}\right|_{\{x_{c}\}} + \mathcal{O}(\delta x_i^3).\nonumber
\end{eqnarray}
This contribution to the free energy can be written $\vec{\delta x^T}F_{ij}\vec{\delta x} \equiv F^{(2)}$, where $F_{ij}$ is the ''fluctuation matrix'' defined by
\begin{equation}
F_{ij}\equiv\langle H_iH_j\rangle - \langle H_i\rangle\langle H_j\rangle=\frac{\partial^2 F}{\partial x_i \partial x_j}.
\label{fluktmatrix}
\end{equation}
This can be diagonalized via an orthogonal transformation by rotating
the original coupling constant basis through an angle $\theta$
\cite{Alonso}, 
providing a new set of coordinates which are linear combinations of coupling constants
\begin{eqnarray}
x_1^\prime &\equiv& x_1\cos\theta +x_2\sin\theta\nonumber\\
x_2^\prime &\equiv& -x_1\sin\theta +x_2\cos\theta,\label{rotation}
\end{eqnarray}
with the corresponding uncorrelated operators 
\begin{eqnarray}
H_1^\prime &\equiv& H_1\cos\theta +H_2\sin\theta\nonumber\\
H_2^\prime &\equiv& -H_1\sin\theta +H_2\cos\theta.
\label{rotatedhamilton}
\end{eqnarray}
This yields
\begin{eqnarray}
F^{(2)} &=& (\delta x_1^\prime, \delta x_2^\prime)\left(\begin{array}{cc}\lambda_+ &0 \\0& \lambda_-\end{array}\right)\left(\begin{array}{c}\delta x_1^\prime\\\delta x_2^\prime\end{array}\right)\nonumber\\ 
&=& (\delta x_1^\prime)^2\lambda_+ +(\delta x_2^\prime)^2\lambda_-, 
\end{eqnarray}
where $\lambda_+$ and $\lambda_-$ are the larger and smaller eigenvalue, respectively. 
If $\lambda_+ \gg \lambda_-$ the first term will dominate the leading order
corrections to $F(\{x_c\})$. 
Hence, to obtain proper scaling, the second derivative of the free energy should be 
evaluated along the direction of the corresponding eigenvector, 
\begin{eqnarray}
\frac{\partial^2 F}{\partial x_1^{\prime 2}} 
\sim |x_1^{\prime}-x_{1c}^{\prime}|^{-\alpha},
\end{eqnarray}
when crossing the critical line.
 
From Eq. (\ref{Model1}), we get the standard expressions
\begin{eqnarray}
\frac{\partial F}{\partial \mu} & = & -\langle H_\mu \rangle \nonumber \\
\frac{\partial^2 F}{\partial \mu^2} & = &  
\langle G_\mu^2 \rangle,  
\label{Mom1+2}
\end{eqnarray}
where $\mu$ is the rotated coordinate along the eigen\-vector corresponding to the largest eigenvalue, and where we have defined $G_\mu = H_\mu - \langle H_\mu \rangle$ with rotated operator $H_\mu$ defined by Eq. (\ref{rotatedhamilton}). The above expressions 
essentially represent generalizations of the expressions for internal energy and 
specific heat. The second derivative contributes to  $C_{\mu}$ which we 
will refer to as the singular part of the second moment of the action, or the 
action susceptibility.  Similarly, we obtain
%\begin{widetext}
\begin{eqnarray}
\frac{\partial^3 F}{\partial \mu^3} =  
\langle G_\mu^3 \rangle,
\label{Mom3}
\end{eqnarray}
%\end{widetext}
and in general we have for the $n$-th moment
 \begin{eqnarray}
\frac{\partial^n F}{\partial \mu^n} =  
\langle G_\mu^n\rangle.
\label{Momn}
\end{eqnarray}

\section{Monte Carlo simulations}

\subsection{Details of the MC simulations}

The critical properties of the models are investigated using large scale 
Monte-Carlo simulations in conjunction with finite-size scaling analysis. A 
Monte-Carlo move is an attempt to replace the field value at one particular 
point by a randomly chosen value. The move is rejected or accepted according 
to the standard Metropolis algorithm. In the case of the XY-model, the phase 
$\theta_j$ is the relevant field, whereas in the full abelian Higgs model both 
the phase and the three components of the gauge field $A_{j,\mu}$ are subjected 
to the Metropolis algorithm. One sweep consists of traversing the system
$L \times L \times L$ while attempting a Metropolis update of each field 
component once.  The acceptance rate of the Metropolis algorithm is kept fixed 
between $60 \%$  and $70 \%$ by dynamically adjusting the maximum allowed changes 
in the fields. However, for the model  defined in Eq. (\ref{Model1}) at large $\beta$,
where $Z_q$ gauge behaviour is expected, the algorithm for controlling the acceptance 
rate has been relaxed by fixing the maximum allowed change to $ 2 \pi$. There is 
no gauge fixing involved in these simulations, and periodic boundary conditions are 
used in all directions. The excess gauge-volume due to the summation in the partition 
function over (redundant) gauge-equivalent field configurations, will cancel out when 
computing all averages of gauge-invariant quantities, in particular when computing 
moments of the gauge-invariant action. 

The third moment of the action will typically behave as shown in Fig. \ref{Fig1}. 
The MC simulations for the compact abelian
Higgs model will be performed for a set of coupling constants that span a line  across 
the phase transition. We diagonalize the ''fluctuation'' matrix Eq. (\ref{fluktmatrix})
and simulate along the trajectory $\beta(\kappa)=\beta_c+a(\kappa-\kappa_c)$, with $a$ 
being determined  from the eigenvector with the largest eigenvalue. 
In all models that we consider in this paper, including benchmarks models,
we then compute the 
moments of the action according to Eq. (\ref{Momn}) applying Ferrenberg-Swendsen 
multihistogram reweighting analysis with jackknifing error estimate. 
For the benchmark models, this procedure produces 
curves as shown in Figs.  \ref{Ising2+3mom} and
\ref{XY2+3mom}. 
The top and bottom values as well as their 
positions for different system sizes are then used to produce scaling plots as shown 
in Fig. \ref{Scalplot}. The combinations of
critical exponents $(1+\alpha)/\nu$ and  $1/\nu$ are then extracted by
bootstrap 
regression analysis.   

\subsection{Simulations of the $8$-vertex model}
\begin{figure}[htbp]
\resizebox{9cm}{13cm}{\includegraphics{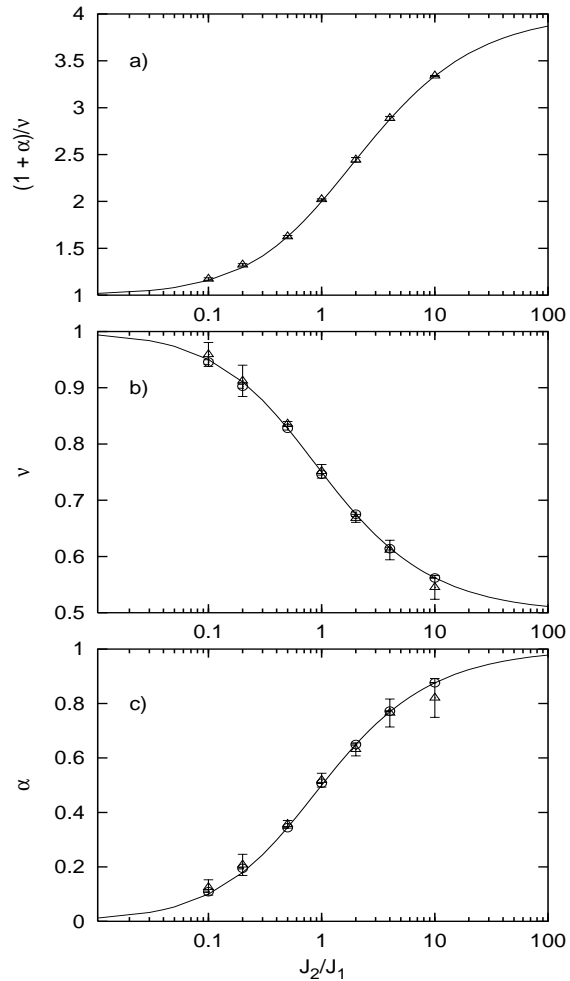} 
\hskip 2cm}
\caption{\label{Expbaxter} a)  $(1+\alpha)/\nu$  from 
FSS finite-size of $M_3$ for Eq. (\ref{Baxter-model}) as a function of 
$J_2/J_1$.  b) Same for the exponent $\nu$, computed directly from
$M_3$ ($\triangle$) and combining results for $(1+\alpha)/\nu$ with 
hyperscaling ($\bigcirc$). c) $\alpha$ as computed directly from $M_3$ 
($\triangle$) and using results for $(1+\alpha)/\nu$ with hyperscaling 
($\bigcirc$). The solid line in b) represent the analytical result 
Eq. \ref{nubaxter}. The solid lines in a) and c) are obtained from
Eq. \ref{nubaxter} and using hyperscaling $ \alpha = 2 - 2 \nu$.
The above results thus also provide a check on hyperscaling
in this model.}
\end{figure}

As a first benchmark on our method of extracting the exponents $\nu$
and $\alpha$ separately, we consider the  $8$-vertex model
\cite{Baxter_book} on a square lattice. This model  has the virtue of
being  exactly solvable, and hence an analytic expression
for the  exponent $\nu$ is known. Moreover, it has  critical exponents 
that are non-universal \cite{Baxter_book}.
The  $8$-vertex model  on a square lattice may be
formulated as a generalized Ising model as follows \cite{Baxter_book}
\begin{eqnarray}
Z_{8V} & = & \sum_{\{ \sigma_i \} } \exp[\beta H_{8V}] \nonumber \\
H_{8V} & = & J_1 \sum_{<<i,j>>} \sigma_i \sigma_j + 
J_2 \sum_{P} \sigma_i \sigma_j \sigma_k \sigma_l
\label{Baxter-model}
\end{eqnarray}
where $\sigma_i = \pm 1$, $<<i,j>>$ denotes a summation over nearest
neighbors (diagonal bonds), and $\sum_P$ denotes a sum over the
elementary plaquettes on the square lattice. 
In the above, we have specialized to the case where spin couplings
along horizontal and vertical bonds  have
been omitted. The critical temperature  and $\nu$ are given by 
\begin{eqnarray}
e^{-2 \beta_c J_2} & = & \sinh(2 \beta_c J_1),  \nonumber \\
\frac{1}{\nu} & = &
2 - \frac{2}{\pi} \cos^{-1} \left( \tanh(2 \beta_c J_2) \right). 
\label{nubaxter}
\end{eqnarray} 
We have computed $M_3$
for this model, using system sizes $L \times L$, with
$L=8,12,20,40,60,80,120,200$. We have used the Metropolis algorithm
with single-spin updates, and up to $5.0 \cdot 10^5$ sweeps over the 
lattice for each coupling constant.
%\begin{figure}[htbp]
%\psfrag{RP}[][][1.7]{$3\cdot10^3$} 
%\psfrag{RP1}[][][1.7]{$2\cdot10^3$} 
%\centerline{\scalebox{0.6}{\rotatebox{0.0}{\includegraphics{2Mom-baxter.ps}}}}
%\caption{\label{Baxter2+3mom} Second moment (upper panel) and third moment
%(lower panel) of the action for the $2D$ Baxter model for system sizes 
%$L=8,12,20,40,60,80,120,200$.}
%\end{figure}
From this, we have extracted scaling plots of the peak-heights and
width between peaks of $M_3$ as a function of system size. 
%the results are shown in 
%Fig. \ref{Scalplot-baxter}
%\begin{figure}[htbp]
%\centerline{\scalebox{0.58}{\rotatebox{0.0}{\includegraphics{Scalplot-baxter.ps}}}}
%\caption{Finite-size scaling plots for the $2D$ Baxter model. a) The peaks in the second
%moment of the action 
%demonstrating that large systems sizes are necessary to obtain correct scaling 
%results. b) The peak to 
%peak values of the third moment of the action, demonstrating that correct scaling 
%results are obtained for much smaller system sizes than what is the case for the 
%second moment. c) The width between peaks in the third moment. }
%\label{Scalplot-baxter}
%\end{figure}
From this, we directly extract the combinations $1/\nu$
and $(1+\alpha)/\nu$, the final results for the exponents are given in
Fig. \ref{Expbaxter}. 
%Note the difference in the quality of the
%scaling plots for the second order moment and the third order moment. 
The agreement between our simulation results and the analytical result 
Eq. \ref{nubaxter} is excellent. 
%Moreover, we see from the scaling
%plots in panels a) and b) of Fig. \ref{Scalplot-baxter} that 
%the third order moment $M_3$ provides superior scaling to the second 
%order moment $M_2$. 
The above demonstrates that for this model, the
third order moment provides an excellent means of extracting non-universal
exponents from  practically accessible system sizes, and in our view 
provides a first excellent and highly non-trivial benchmark test on
the method. We next provide further benchmark tests on a number of $3D$
systems, before going on to extracting exponents for the
$2+1$-dimensional compact abelian Higgs model.

\subsection{Simulations of the $3D$ XY-, Ising-, and Ising ($Z_2$) gauge-models}
We reemphasize that what follows in this section is a benchmark on the method of 
bringing out non-analytic thermodynamics without recourse to order parameter 
measurements, by measuring third moments of the action. In later sections the method 
will be applied to a lattice gauge model for which no non-local order parameter 
currently is known, and where we will obtain precise values of critical exponents. In 
the present section, we will reproduce known results for the critical exponents $\alpha$ 
and $\nu$ for the $3D$XY model, and the two $3D$ Ising models, directly from third 
moments of the action, thus benchmarking the method. We will demonstrate that the method 
of using the third moment of the action is far superior to using the second moment, for 
practical system sizes accesible in numerical simulations. This is due to the fact that 
the third moment is not contaminated by contributions from the second moment, which 
unfortunately often is marred by contributions from confluent, or subdominant, singularities. 
   
The $XY$ and Ising models are defined by the following two partition functions, 
\begin{eqnarray}
Z_{\rm{XY}}&=&\int_{-\pi}^\pi\left[\prod_{j=1}^{N} \frac{d \theta_j}{2\pi}\right]  
\exp (\beta H_{\rm{XY}} ), \nonumber \\
H_{\rm{XY}}&=&\sum_{{\rm j}, \mu} \cos(\Delta_{\mu} \theta_j) , \nonumber \\
Z_{\rm{I}}&=& \sum_{ \{ \sigma_i = \pm 1\} }  \exp(\beta H_{\rm{I}}), \nonumber \\
H_{\rm{I}} &=& \sum_{\langle i,j \rangle} \sigma_i \sigma_j. \nonumber 
\end{eqnarray}
In addition, we consider the three-dimensional Ising lattice gauge 
theory \cite{Wegner} 
\begin{eqnarray}
Z_{\rm{IGT}} & = & \sum_{ \{ U_{ij} = \pm 1\} }  \exp (\beta H_{\rm{IGT}}), \nonumber \\
H_{\rm{IGT}} & = & \sum_{\rm{P}}  U_{i j} U_{j l} U_{l k} U_{k i}, \nonumber 
\label{Models}
\end{eqnarray}
where the summation in the last expression runs over all plaquettes $P$ of the 
lattice. We note for later use that the Ising model ($Z_2$ spin model) defined 
by $Z_I$ and the Ising gauge theory ($Z_2$ gauge theory) defined by $Z_{IGT}$ 
are dual to each other in $d=3$ \cite{Savit1}.

The second and third moments $M_2$ and $M_3$ we will consider for these models are 
given by
\begin{eqnarray}
M_2=\frac{\partial^2 F}{\partial \beta^2} &=& \langle G_\lambda^2 \rangle, \nonumber \\
M_3=\frac{\partial^3 F}{\partial \beta^3}&=& \langle G_\lambda^3 \rangle,
\label{Moments}
\end{eqnarray}
where $\lambda \in ({\rm{XY}}, \rm{I}, \rm{IGT})$, and where $\rm{I}$ and $\rm{IGT}$ 
denote the Ising- and Ising gauge theories, respectively.
In all these cases, scaling $\sim L^{\alpha/\nu}$ is expected for
the peak in the second moment, while scaling $\sim L^{(1+\alpha)/\nu}$
and $\sim L^{-1/\nu}$ is expected for the third moment peak-to-peak
height and width respectively as indicated in Fig. \ref{Fig1}. We mention in 
passing that for the $3D$XY model, we have $(1+\alpha)/\nu=1.467$, while 
$(1+\alpha)/\nu=1.763$ for the $3D$ Ising model and $3D Z_2$ 
lattice gauge theory. The latter follows from the fact that
models which are connected by duality transformations have 
identical values of $\alpha, \nu$, since these two exponents 
can be obtained from scaling of the free energy.
On the other hand, {\it certain combinations} of the remaining 
critical exponents, which depend on the degrees of freedom one 
chooses to describe the transition with, remain invariant. These 
invariant combinations are given by \cite{Hovethesis}
\begin{eqnarray}
 \frac{\gamma}{2 - \eta}, ~~ 2 \beta + \gamma, ~~ \beta(\delta+1).
\end{eqnarray}
The first is a consequence of Fisher's scaling law, the other two follow 
from Rushbrooke´s scaling law.

\begin{figure}[htbp]
%\psfrag{RP}[][][1.7]{$3\cdot10^3$} 
%\psfrag{RP1}[][][1.7]{$2\cdot10^3$} 
\centerline{\scalebox{0.6}{\rotatebox{0.0}{\includegraphics{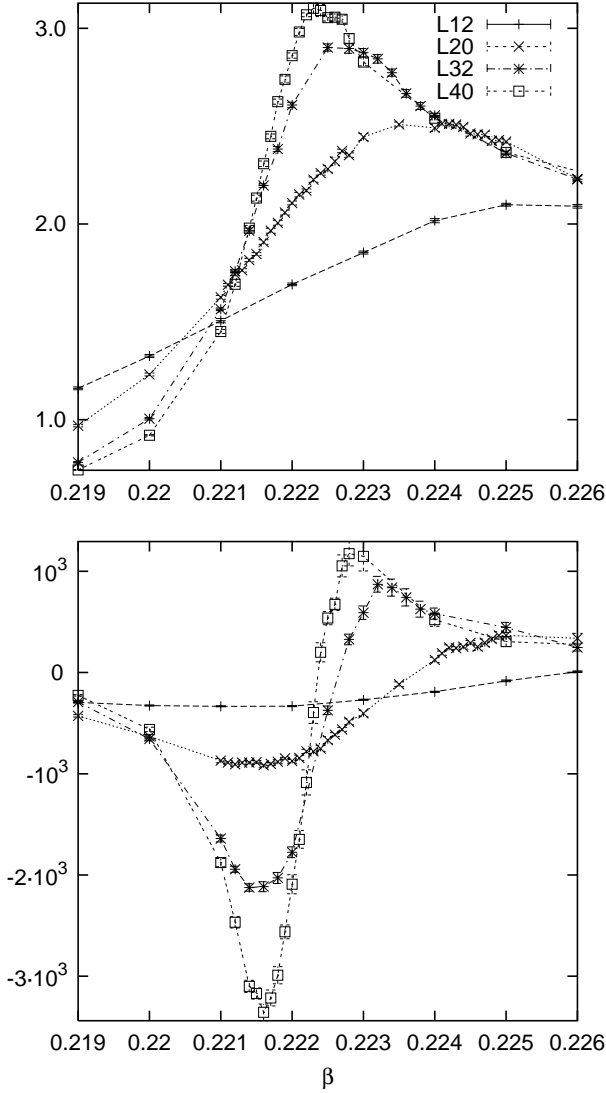}}}}
\caption{\label{Ising2+3mom} Second moment (upper panel) and third moment
(lower panel) of the action for the $3D$ Ising model for system sizes 
$L=12,20,32,40$.}
\end{figure}

\begin{figure}[htbp]
\centerline{\scalebox{0.6}{\rotatebox{0.0}{\includegraphics{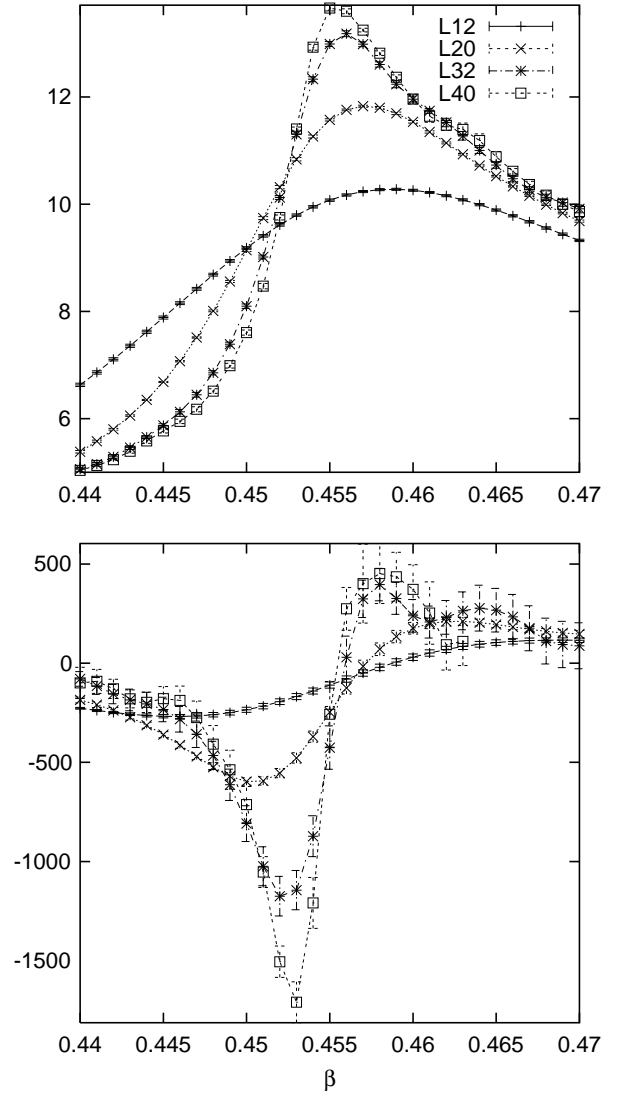}}}}
\caption{\label{XY2+3mom} Second moment (upper panel) and third moment
(lower panel) of the action for the 
$3DXY$ model for system sizes $L=12,20,32,40$.}
\end{figure}

Fig. \ref{Ising2+3mom} shows the second and third moments for the $3D$ 
Ising model for system sizes $L=12,20,32,40$ as a function of $\beta$, 
while Fig. \ref{XY2+3mom} shows the corresponding quantities for the 
$3D$XY model.

\subsection{Finite-size scaling of $M_2$ and $M_3$}
\begin{figure}[htbp]
\centerline{\scalebox{0.58}{\rotatebox{0.0}{\includegraphics{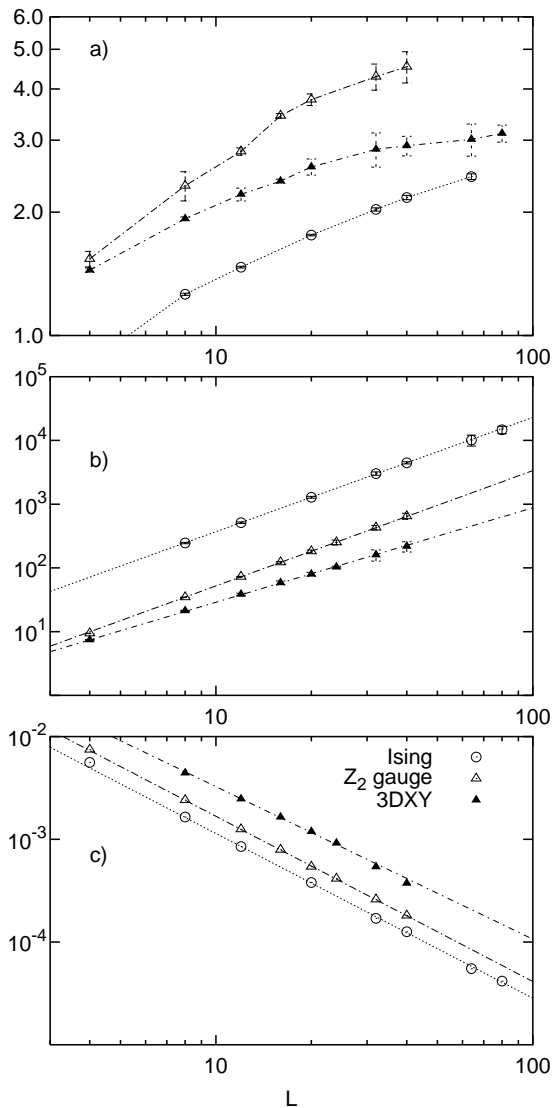}}}}
\caption{Finite-size scaling plots for the $3D$XY, $3D$ Ising, and $3D$ $Z_2$ lattice 
gauge model defined in Eq. \ref{Models}. a) The peaks in the second
moment of the action 
demonstrating that very large systems sizes are necessary to obtain correct scaling 
results. For clarity, two of the plots are lowered with a multiplicative factor of 0.7 
and 0.45 for the Ising- and Ising lattice gauge models, respectively. b) The peak to 
peak values of the third moment of the action, demonstrating that correct scaling 
results are obtained for much smaller system sizes than what is the case for the 
second moment. Note also the close similarity of the scaling results for the Ising 
and $Z_2$ lattice gauge models, which should be indentical, up to
constants, due to duality. This 
serves as a quality check on the simulations. c) The width between peaks in the 
third moment. }
\label{Scalplot}
\end{figure}

We first consider the second moments of the $3DXY$ model and the $3D$ 
Ising model. These exhibit peaks at $T_c$ which in principle are 
amenable to finite size scaling. Fig. \ref{Scalplot} upper panel shows 
the finite-size scaling plots of the peaks in the second moment of both models. 
As is seen from the figure, the peaks grow as $L$ increases. However, it is 
clear that none of the scaling plots have reached their asymptotic behaviour.

For the $3D$XY model, the apparent scaling for small to intermediate values 
of $L$ is clearly spurious, as it levels off for large system sizes. In 
principle, a {\it negative } slope should eventually be obtained for 
asymptotically large system sizes, but the general experience is 
that impractically large system sizes are required to see this, let 
alone estimate the slightly negative value of $\alpha$ with any precision.  

For the $3D$ Ising model, the situation is different in that the 
peak height grows steadily as $L$ increases, eventually approaching 
a  straight line on a double-logarithmic plot. Although the quality 
of the scaling improves with increasing $L$, an approximate evaluation
of the slope yields $\alpha/\nu=0.3$ whereas the known value is
$\alpha/\nu=0.175$. Hence, it is clear that inaccesibly large systems 
are required to obtain the specific heat exponent with any
accuracy from the second moment scaling analysis.

%\begin{figure}[h,t]
%\resizebox{17cm}{!}{\includegraphics{2Mom.scal.plot_ising_xy.ps} 
%\hskip 2cm \includegraphics{2Mom.scal.plot_ising_xy.ps}}
%\caption{Finite-size scaling of the peak in the second moment of 
%the action for the $3D$XY and $3D$ Ising models, demonstrating
%that very large systems sizes are necessary for obtaining correct
%scaling results. } 
%\label{2momscal}
%\end{figure}

%
%
%
%\subsection{Finite-size scaling of $M_3$}
%
%\begin{figure}[htbp]
%\centerline{\scalebox{0.60}{\rotatebox{0.0}{\includegraphics{3Mom.scal.plot_ising_xy.ps}}}}
%\caption{\label{3momscal}Finite-size scaling of the peaks in the third moment of 
%the action for the $3D$XY, $3D$ Ising, and $3D$
%$Z_2$ lattice gauge model defined in Eq. \ref{Models}, 
%demonstrating that correct scaling results are obtainedd for 
%much smaller  system sizes than what is the case for the 
%second moment. Note also the closeness of the scaling results
%for the Ising and $Z_2$ lattice gauge models, which should be
%indentical due to duality. This serves as a quality check
%on the simulations. }
%
%\end{figure}
%\begin{figure}[htbp]
%\centerline{\scalebox{0.60}{\rotatebox{0.0}{\includegraphics{3Mom.scal.with_ising_xy.ps}}}}
%\caption{\label{widthscal}Finite-size scaling of the width between peaks in the third moment of 
%the action for the $3D$XY, $3D$ Ising, and $3D$
%$Z_2$ lattice gauge model defined in Eq. \ref{Models}.}
%\end{figure}

Consider now the third moment of the action for the $3D$XY and Ising
models. The results are given in Fig. \ref{Scalplot}, middle panel. It is  
obvious that the quality of the scaling in both cases is vastly improved 
compared to the results obtained 
from the second moment. This demonstrates rather convincingly that 
finite-size scaling of the third moment of the action, i.e. a purely 
thermodynamic  measurement without recourse to
order parameters, suffices to bring out that a singular part of the free 
energy exists in both cases. This  would have been difficult to 
conclude  based on the second moment of the action, at least for the 
$3D$XY model. Note that scaling such as this would not have been found 
had there not been a non-analytic part of the free energy. Thus, 
the quality of the scaling alone suffices to demonstrate the existence 
of a phase-transition. Moreover, the double-peak structure in $M_3$
permits a  separation of $\alpha$ and $\nu$ without recourse to
hyperscaling. The results for the widths between the peaks in the
histrograms are given in Fig. \ref{Scalplot}, bottom 
panel.  Note the good quality of the scaling.

\begin{table}[h]
\label{Sweeps}
%\begin{ruledtabular}
\caption{\label{tabell}Critical exponents extracted for the benchmark
models, where $\alpha$ is computed by combining the $(1+\alpha)/\nu$
and $1/\nu$ results.}
\begin{tabular}{lccc}
\hline
Model         & $(1+\alpha)/\nu$         &$\nu$                 &   $\alpha$         \\
\hline
XY            &   1.46   $\pm$0.01       & 0.67  $\pm$0.01      & -0.01   $\pm$0.01  \\ 
Ising         &   1.77   $\pm$0.01       & 0.63  $\pm$0.01      &  0.11   $\pm$0.01  \\ 
$Z_2$ Gauge   &   1.78   $\pm$0.02       & 0.63  $\pm$0.01      &  0.12   $\pm$0.02  \\
\hline
\end{tabular} 
%\end{ruledtabular}
\end{table}
%\begin{table}[h]
%\label{Sweeps}
%%\begin{ruledtabular}
%\caption{\label{tabell}Critical exponents extracted for the benchmark
%models, where $\alpha$ is computed by combining the $(1+\alpha)/\nu$
%and $1/\nu$ results.}
%\begin{tabular}{lccc}
%\hline
%Model         & $(1+\alpha)/\nu$           &$\nu$                   &   $\alpha$           \\
%\hline
%XY            &   1.464   $\pm$0.012       & 0.673  $\pm$0.007      & -0.014   $\pm$0.013  \\ 
%Ising         &   1.768   $\pm$0.012       & 0.628  $\pm$0.003      &  0.111   $\pm$0.008  \\ 
%$Z_2$ Gauge   &   1.783   $\pm$0.019       & 0.630  $\pm$0.009      &  0.123   $\pm$0.024  \\
%\hline
%\end{tabular} 
%%\end{ruledtabular}
%\end{table}

The benchmark results for the three models are given in Table
\ref{tabell}. We see that the error bars of these benchmark values all
include the known values of $\alpha$ and $\nu$. Finally,
the exponents obtained for the $3D$ Ising spin model and
the $3D$ Ising gauge theory,
are within error bars found to be the same, as they should be from duality of
the two theories in three dimensions. Since the degrees of freedom used in
the simulations are vastly different, this serves as a highly nontrivial 
quality check on the simulations. 

\subsection{Benchmark simulations of a 3DXY model coupled to a $Z_2$ clock model}

To investigate the scaling behaviour of the third moment of the action in a
well studied $3D$ theory \cite{Pelcovits} 
with two coupling constants and two fixed points
of type $Z_2$ and $XY$, we have considered the model
\begin{eqnarray}
\label{h3dxy}
Z & = & \int_{-\pi}^\pi\left[\prod_{j=1}^{N} \frac{d \theta_j}{2\pi}
\right]
 \exp \left[\beta H_{XY} +  hH_{Zq} \right] \nonumber \\
H_{XY} &=& \sum_{j, \mu}\cos(\nabla_\mu\theta_j)  \nonumber \\
H_{Zq} &=& \sum_j \cos(q\theta_j),
\end{eqnarray}
defined on a $d=3$ dimensional cubic lattice with $N$ sites. The phase
$\theta_j$ resides on every site $j$ and $q$ is integer
valued. 

Let us consider the $q=2$ theory. By simple inspection of the model we find 
that the limit $h\rightarrow\infty$ leads to a twofold symmetry constraint 
on the phase, and the model becomes exactly the $3D$ Ising model. The limits
$\beta\rightarrow\infty$ and $\beta = 0$ are trivial
theories. From renormalization group theory, the $q=2$ model is found
to have a $XY$ phase transition with $h = 0$, while it exhibits Ising
exponents for any other $h\neq 0$(at least in the vicinity of $h =
0$.) 

\begin{figure}[h,t]
\centerline{\scalebox{0.60}{\rotatebox{0.0}{\includegraphics{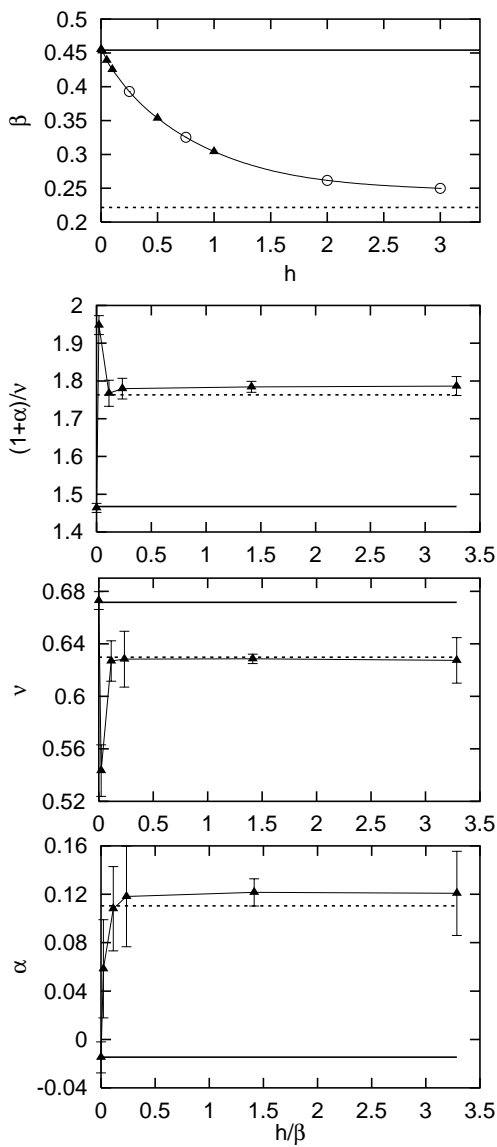}}}}
\caption{\label{h_plots} Upper panel: The phase diagram of the model
Eq. \ref{h3dxy}. The open circles denote points where only the
phase transition has been located, while the filled triangles denote
points where also the critical exponents have been measured.
Lower panels: The three lower panels show the combination of
exponents $(1+\alpha)/\nu$, as well as $\nu$ and $\alpha$, as a
function of $h/\beta$. Note the  rapid crossover in  $\alpha$ from
the $3DXY$ value at $h = 0$ to the 3D Ising value as $h$ is increased.
The dotted and horizontal lines denote $3DXY$ and $3D$ Ising values,
respectively, for the various quantities.} 
\end{figure}

We have performed Monte Carlo simulations on the model
Eq. (\ref{h3dxy}), $q = 2$ with system sizes up to $L =32$. The
resulting exponents from the third moment scaling analysis are given
in Fig. \ref{h_plots}. We measure $Z_2$ critical exponents down
to very  small values of $h$. However, it is clear that the critical
coupling approaches the $3D$XY value as shown in the upper panel.
Note the dramatic increase in both $(1+\alpha)/\nu$ and $1/\nu$ for 
$h = 0.01$ in the vicinity of the $XY$ fixed point, where we find 
that the exponents become even less $XY$ like. The value for 
$\alpha$ shows a more smooth crossover behavior from 3D$XY$ value to
3D Ising value as $h$ is increased from $0$. However, the scaling 
appears not to have reached its asymptotic behaviour and the slopes 
are clearly decreasing for increasing system sizes. We expect this 
to be due to a cross-over regime in which the cross-over exponent 
contaminates our estimates, but that all the combinations
of exponents eventually will converge towards Ising values,
when $h \neq 0$.

%Hence, FSS analysis with third moment of the action in the vicinity
%of a fixed point might be marred by corrections to scaling and may
%require simulations on larger system sizes. 

%Previously we have considered the posibility that the $3D$ compact Abelian Higgs
%model with $q = 2$ is a theory with \textit{two} fixed points. By
%comparing Fig. \ref{h_plots} with Fig. \ref{} we find a similar
%peak behaviour in the critical exponents where they leave the $Z_2$
%scheme. Hence the model Eq. \ref{Model1} with $q = 2$ might have an attractive
%$Z_2$ fixed \textit{point}. (perhaps followed by a non-universal area
%on the critical line...?)

\subsection{Simulations of the $3D$ compact abelian Higgs model, $q=2$}

We next consider the third moment of the model defined by Eq. (\ref{Model1}). 
The model with $q=2$ has previously been studied by Monte Carlo simulations on 
small systems $L^3$, with $L=2,3,4$  many years ago \cite{Bhanot} using second
moments of the action. A short version of the results to be presented in this 
section, has already appeared \cite{sudbo1}. To our knowledge, no simulations 
have been performed in $d=3$ between those presented in Ref. \onlinecite{Bhanot} and 
the much more recent large-scale simulations that have
been performed \cite{sudbo1}. Due to 
our observation that scaling of the second moment 
requires much larger systems than those studied in \cite{Bhanot} in order to 
obtain reliable results for exponents, a revisit to the problem seems very 
appropriate. This is particularly true given the importance the model now has 
acquired in condensed matter system as a real model for strongly correlated 
quantum systems at zero temperature  in two spatial dimensions
\cite{RS,Wen,Senthil,Motrunic}. In Ref. \onlinecite{Motrunic}, precisely such 
an effective model with $q=2$ was derived from a proposed microscopic description 
of {\it charge-fractionalized} phases in strongly correlated systems.  
Specifically, the confined phase of the $q=2$ model was interpreted as a 
Mott-Hubbard insulating phase, while the deconfined-Higgs phase was interpreted 
as a charge-fractionalized insulating phase.

\begin{figure}[htbp]
\resizebox{9cm}{17cm}{\includegraphics{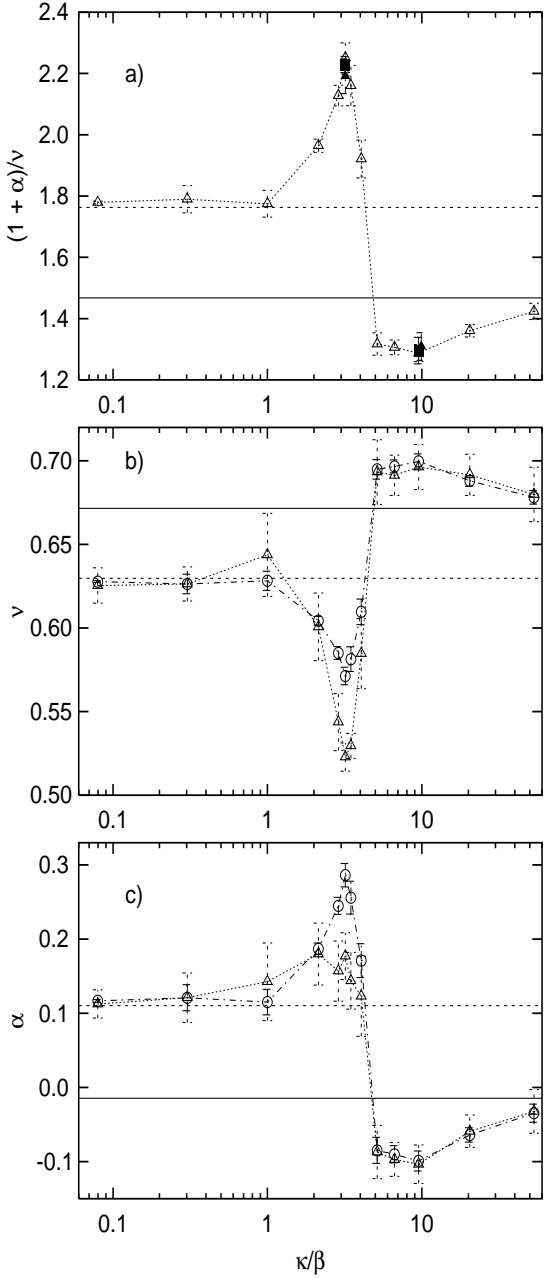} 
\hskip 2cm}
\caption{\label{Expq2} a)  $(1+\alpha)/\nu$  from 
FSS finite-size of $M_3$ for Eq. (\ref{Model1}) for $q=2$. Note the
variation relative to the $Z_2$-limit $1.76$ (dotted horizontal line)  
and the $U(1)$-limit $1.467$ (solid horizontal line). b) Same for the 
exponent $\nu$, computed directly from $M_3$ ($\triangle$) and combining 
results for $(1+\alpha)/\nu$ with hyperscaling ($\bigcirc$). 
c) $\alpha$ as computed directly from $M_3$ 
($\triangle$) and using results for $(1+\alpha)/\nu$ with hyperscaling 
($\bigcirc$). The maximum and minimum in a) have been obtained by crossing 
the critical line along the trajectory $\beta(\kappa) = \beta_c + 
a ~ (\kappa - \kappa_c)$ with $a = \infty$ ($\triangle$), $a = 1$ 
($\blacksquare)$, and $a=-1$ ($\blacktriangle$) using $\beta_c = 
0.665, \kappa_c = 2.125$ (max.), and $\beta_c = 0.525, \kappa_c = 
5.0$ (min.).}
\end{figure}

The critical exponents obtained from third moment FSS analysis are presented
in Fig. \ref{Expq2}. The exponents $\alpha$, $\nu$, and the combination 
$(1+\alpha)/\nu$ vary continuously along the critical line showing $3D$XY and 
$Z_2$ universality in the $\kappa\to\infty$ and $\beta\to \infty$ limits, 
respectively. The $Z_2$ behaviour persists deep into the phase diagram, 
while we find a broad
nonuniversal area in the large-$\kappa$ region. These two areas are
joined by a peak in the exponents. To check the dependence of the
trajectory, we performed simulations for several slopes at the two
extremas, using $a=\infty$, $a=1$, $a=-1$ and the direction given by
the diagonalization of the fluctuation matrix. Within error bars, all
slopes consistently produced the same exponents.

The results appear to rule out that  
$Z_2$- and $XY$-critical behaviors are isolated points at the extreme 
ends of the critical line. However, a feasible suggestion could be that two types of
universality, $Z_2$ and $XY$, are separated at a multicritical point on the
critical line. We believe 
this to be ruled out by the strong deviation in $(1+\alpha)/\nu$ from $Z_2$- 
and $XY$-values at intermediate $\kappa/\beta$.  On balance, we thus conclude  
that the model Eq. (\ref{Model1}) defines a {\it fixed-line theory}, rather 
than exhibiting two scaling regimes separated by a multicritical point. However, 
the $Z_2$ character of the confinement-deconfinement transition persists  to 
surprisingly large values of $\kappa/\beta$ on the critical line, see
also Fig. 5 of Ref. \onlinecite{Bhanot}. Fixed-line theories in $2+1$ dimensions 
are known \cite{Note3}, and non-universal exponents imply the existence of 
marginal operators  in Eq. (\ref{Model1}), yet to be identified.

%This shows that the critical line for 
%the model Eq. (\ref{Model1}) is not a fixed point in the renormalization group
%sense, since this would have yielded universal exponents, independent
%of $(\beta,\kappa)$. Moreover, it does not represent merely {\it two} fixed points, 
%namely $Z_2$ and $XY$, since this would have yielded $(1+\alpha)/\nu = 1.76$ or 
%$1.467$ in two regions of the  critical line above and below some finite value of 
%$\kappa$, and otherwise independent of $(\beta,\kappa)$. We conclude that the 
%critical line  $\beta_c(\kappa)$ is a {\it fixed line theory} with continuously 
%varying  exponents depending on $(\beta,\kappa)$.

\subsection{Simulations of the $3D$ compact abelian Higgs model,
$q=3$}

The $Z_q$ spin model and $q$-state Potts model are easily seen to be 
equivalent for $q=2$ and $q=3$ \cite{Note2}. For the $q$-state
Potts model, it is known that when one 
generalizes $q$ to be real-valued, the phase transition in the model 
in $d=3$ changes from continuous to discontinuous when $q$ is increased
beyond the value $q=2.625$ \cite{Gliozzi}. Hence, both the $Z_3$ lattice 
gauge theory and the $Z_3$ spin model have first order
phase transitions. In the limits $\beta\rightarrow\infty$ and
$\kappa\rightarrow\infty$ the model defined by (\ref{Model1}) reduces
to the $Z_3$-spin model and 3DXY model respectively. If these
phase transitions survive on the critical line for finite coupling values, 
 a tricritical point joining the first and second order critical lines is 
expected to exist. As we shall see, this is precisely what happens.

\begin{figure}[h,t]
\centerline{\scalebox{0.5}{\rotatebox{0.0}{\includegraphics{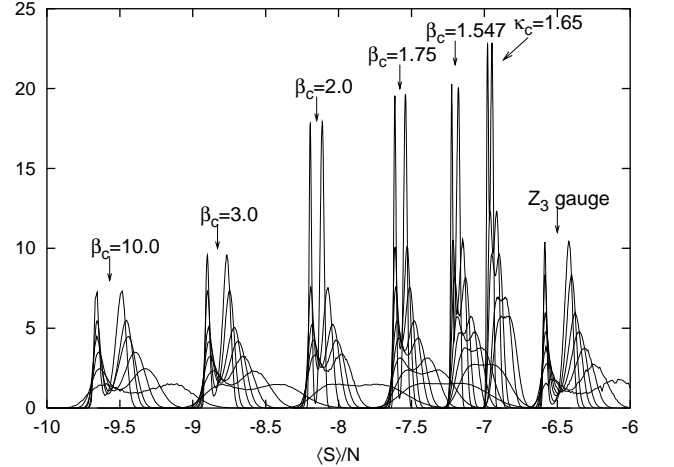}}}}
\caption{Normalized histograms of the action (\ref{Model1}) for $q=3$
at the first order phase transition with various critical couplings
and system sizes $L = 8, 12, 16, 20, 24, 32, 40, 64$. The histogram 
height increases with system size. The horizontal axis shows the
average value of the action per unit volume, and the histograms have 
been shifted horizontally for clarity.} 
\label{histograms}
\end{figure}

Therefore, we reach the important 
conclusion that the model exhibits a first order phase-transition not 
only for $\beta = \infty$, Eq. (\ref{Theory}), but also for {\it finite 
values of $\beta$}. To identify and investigate the first order phase 
transitions we have computed histograms of the action (\ref{Model1}) 
at critical coupling constants. These have a 
double-peak structure due to the two coexistiting phases that
characterize a first-order phase transition. The
histograms have been produced from large scale Monte-Carlo 
simulations followed by muliti-histogram reweighting and
jackknife error analysis \cite{Hove1}. The results are shown 
in Fig. \ref{histograms}. 

\begin{figure}[h,t]
\centerline{\scalebox{0.5}{\rotatebox{0.0}{\includegraphics{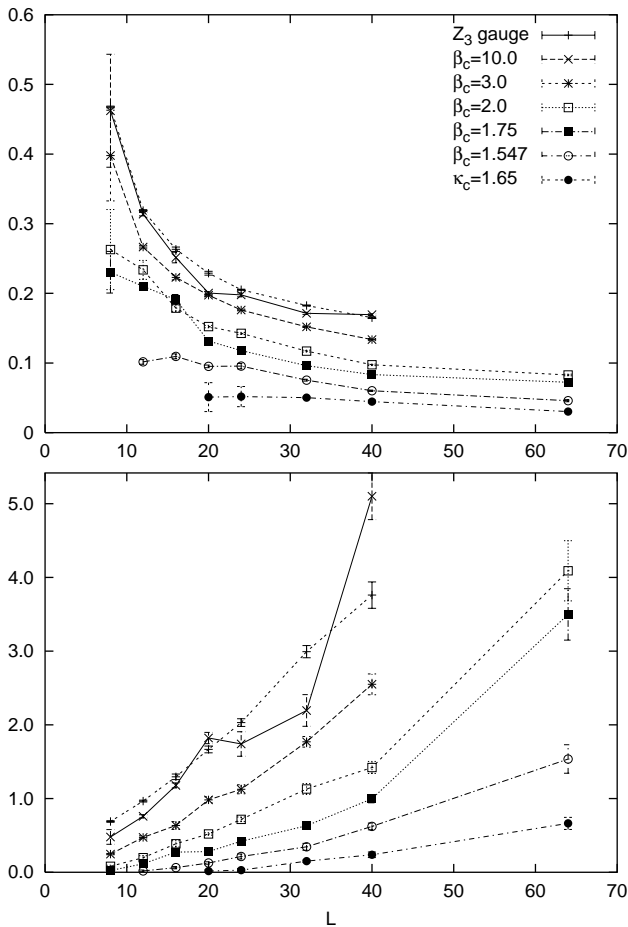}}}}
\caption{Upper panel: The discontinuity of the action (\ref{Model1}) as a
function of system size. Note that the plots seem to be converging
towards a finite energy value. Lower panel: The energy difference
between the two coexisting phases $\Delta F(L)$ as a function of
system size, same labeling as in the upper panel.} 
\label{latentheat}
\end{figure}

A first-order phase transition is characterized by two coexisting
phases with the same free energy. There should therefore exist domain
walls separating the two phases. The area of the domain walls is related 
to the energy difference $\Delta F(L)$ required to keep the two phases 
separated through the expression
\begin{equation}
\Delta F(L) = \ln P(S,L)_{\max} - \ln P(S,L)_{\min} \sim L^{d-1},
\end{equation}
where $P(S,L)$ is the probability for a value $S$ of the action in a
system of size $L^d$, and $L^{d-1}$ is the cross-section area between
the ordered and disordered phases\cite{LeeKosterlitz}. The results 
shown in Fig.\ref{latentheat} confirm this for system sizes $L \ge 24 $.

The discontinuity of the action or equivalently the width between the 
peaks in the histograms is, strictly speaking, not the latent heat. Nonetheless, 
it can be taken as a reasonably accurate measure of this quantity along a 
sufficiently small part of the phase transtion line. It is important to 
ascertain whether or not the first-order character of systems persists in 
the thermodynamic limit. Hence, we plot the discontinuity of the action as a 
function of system size in Fig. \ref{latentheat}, upper panel, and find that 
the discontinuity of the action is finite at least up to $\kappa_c=1.65$. The 
first order phase transition becomes increasingly weaker, i.e. the latent heat 
in the transition is reduced as we approach $\kappa_c = 1.65$ from below.

%On the critical line with $\kappa_c \ge 1.70$ we see no sign of a
%first-order phase transition. Since we measeure 3DXY critical exponents
%in most of this region, we conclude the existence of a
%\textit{tricritical point} somewhere in the region of $1.65 <
%\kappa_{\rm{tri}} < 1.70$. 

\begin{figure}[h,t]
\centerline{\scalebox{0.60}{\rotatebox{-90.0}{\includegraphics{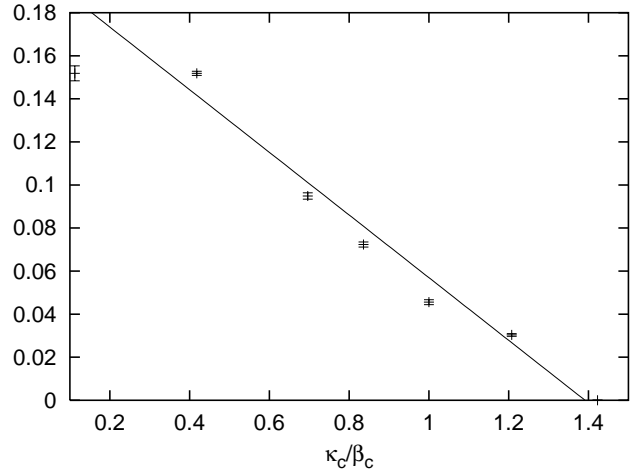}}}}
\caption{Width of peaks in histograms as a function of $\kappa_c/\beta_c$. The solid
line is a linear fit to the data points for $\kappa_c/\beta_c > 0.8$.} 
\label{tricrit}
\end{figure}

Approaching the tricritical point along the first order line, the discontinuity
in the action, equivalently the width between the peaks in the histograms, must 
vanish. From Fig. \ref{latentheat} we have extrapolated the thermodynamic limit 
of  this quantity, and plotted them as a function of the critical coupling in
Fig. \ref{tricrit}. A linear extrapolation yields an
estimate for the tricritical point $\kappa_{\rm{tri}}/\beta_{\rm{tri}}
= 1.39 \pm 0.06$ corresponding to 
$(\beta_{\rm{tri}}, \kappa_{\rm{tri}})=(1.23 \pm 0,03,1.73 \pm 0.03)$.

\begin{figure}[h,t]
\resizebox{9cm}{17cm}{\includegraphics{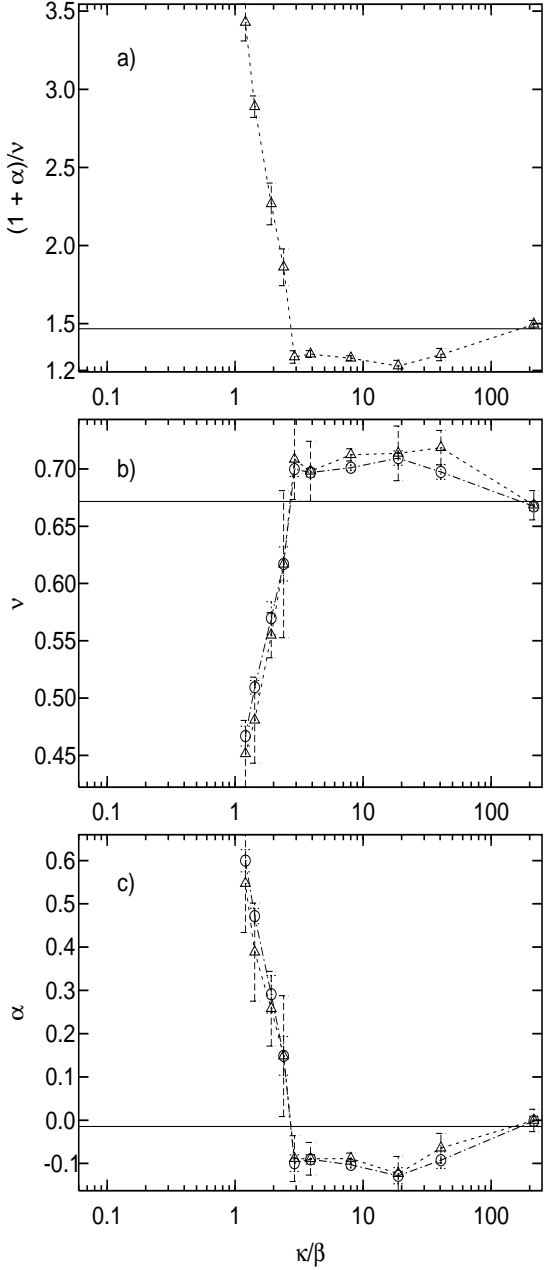} 
\hskip 2cm}
\caption{ a)  $(1+\alpha)/\nu$  from 
FSS finite-size of $M_3$ for Eq. (\ref{Model1}) for $q=3$. Note the
variation relative to the $U(1)$-limit $1.467$ (solid horizontal line) and 
the violent behaviour near the critical point $\kappa_{tri}$. b) Same for the 
exponent $\nu$, computed directly from $M_3$ ($\triangle$) and combining 
results for$(1+\alpha)/\nu$ with hyperscaling ($\bigcirc$). c) $\alpha$ as 
computed directly from $M_3$ ($\triangle$) and combining results for 
$(1+\alpha)/\nu$ with hyperscaling ($\bigcirc$). 
\label{q=3exponents}}
\end{figure}

On the other side of the tricritical point we have performed a third moment
analysis. The exponents are expected to be those of $3D$XY in the large-$\kappa$ 
limit. Strictly speaking, critical exponents are  properties of second order phase 
transitions. For a first order transition the correlation length stays finite 
and hence there is no scale-invariance. However, bounds for some of the 
exponents can be obtained and the limits of these bounds correspond to the
exponents one would get by relaxing the definitions and formally considering 
first order transitions\cite{Hovethesis}. One finds that the limits are 
$\alpha=1$ and $\nu=1/d$ respectivly. Hence when approaching the tricritical 
point we expect $(1+\alpha)/\nu=6$ and $\nu=1/3$.

The critical exponents, given in Fig. \ref{q=3exponents}, agree with the
expectations. The exponents are XY like in the large $\kappa$ limit while in 
the region $2.4\lesssim\kappa\lesssim 50$ the quantity $(1+\alpha)/\nu$ is
lower than the XY value and hence exhibits nonuniversiality similar to what 
we found for $q=2$. At approximately $\kappa\sim 2.4$ the exponents rise 
abruptly towards the first order limiting values. Due to super-critical 
slowing down, obtaining good quality third moment scaling plots becomes 
difficult when approaching the tricritical point. Thus, locating the critical 
point by deciding where the exponents have reached the values one expect
in the {\it limit} of a first order phase transition, is not feasible.

For $q=3$, which is a special case in this context (see below), this is very much 
like what is now known to happen in the non-compact abelian Higgs model in $d=3$.
In that system, a first order phase transition characteristic of type-I superconductivity 
at small values of the Ginzburg-Landau parameter, is converted to a 
second order phase transition characteristic of type-II superconductivity.
The tricritical value of the Ginzburg-Landau parameter where this
change in the character of the phase transition occurs, has recently
been determined with precision in large-scale Monte Carlo simulations, to 
be given by $0.8/\sqrt{2}$ \cite{Mo}. This is in remarkable agreement 
with previous analytical results using duality arguments \cite{Kleinert}. 

\subsection{The phase diagram for $q=2,3,4$}
\begin{figure}[htbp]
\centerline{\scalebox{0.35}{\rotatebox{270.0}{\includegraphics{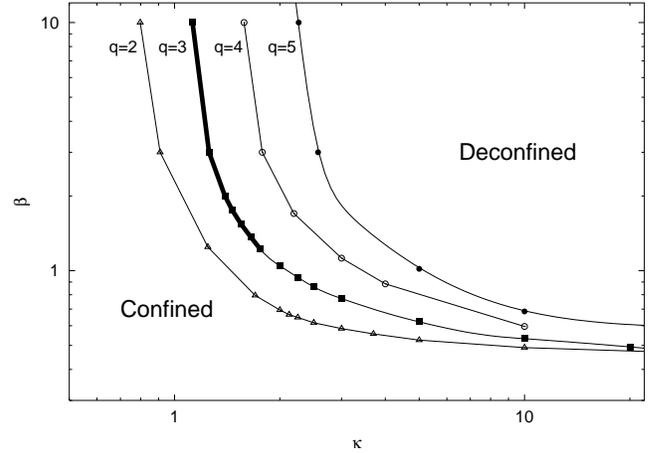}}}}
\caption{\label{phaseq=2} 
The phase diagram for the $d=3$ compact abelian Higgs model in three 
dimensions for $q=2,3,4,5$. All lines are critical  for all values of $\kappa$ 
except for the case $q=3$, which is first order for $\kappa < \kappa_{\rm{tri}}$
and second order otherwise. The thick solid portion of the $q=3$-line denotes 
a first-order transition. Detailed finite-size scaling analysis of the line
$q=2$ shows that the critical exponents $\alpha$ and $\nu$ vary continuously
along the critical line. This makes the theory defined by Eq. (\ref{Model1}) 
a {\it fixed-line} theory with non-universal critical exponents, as opposed
to a fixed-point theory with universal exponents. The critical line approaches
the $3DXY$ value $\beta_c=0.453$ as $\kappa \to \infty$ for all values of 
$q \in \mathbb{N}$.}
\end{figure}

Let us summarize what has been discussed above.
The phase structure of the model is given in Fig. \ref{phaseq=2}. 
From the early  Monte-Carlo simulations on the model \cite{Bhanot} for $q=2$, 
it  is known that the critical  line $\beta_c(\kappa)$  approaches 
$\lim_{\kappa \to \infty} \beta_c = 0.454$ while  when $\beta \to \infty$, 
there is a critical value of $\kappa$ given by $\kappa_c = 0.761$. As $q$ 
increases, the vertical part of the phase-transition line moves up in 
$\kappa$, while for the values of $q$ we have considered, the lines  are 
critical except for the case  $q=3$ which is a first order line for large 
values of $\beta$. For {\it any} $q$, however, we know that the limit $\kappa \to \infty$ 
represents the $U(1)$ limit, which must exhibit a second order 
phase transition. Hence, the line for $q=3$ in Fig. \ref{phaseq=2} must 
contain a {\it tricritical} point.  The first order line for large $\beta$ 
terminates at a tricritical point, and is second order for larger values of 
$\kappa$. For $q=4$, which is also shown, the phase-transition line is second 
order. The limit £$\beta \to \infty$ corresponds to the $d=3$ $Z_4$ lattice 
model which is dual to the $Z_4$ spin model. Using the fact that the symmetry 
group $Z_4 = Z_2 \otimes Z_2$, the universality class  of the $q=4$ phase 
transition line is expected to interpolate between the Ising case in the 
limit $\beta \to \infty$ and the $3dXY$ universality class in the limit 
$\kappa \to \infty$.

Based on what is presented above, we conclude that for $q \neq 3$, the
model Eq. (\ref{Model1}) exhibits a critical line $\beta_c(\kappa)$ for all
values of $\kappa$, while for  $q = 3 $ the phase transition of the
model is first order for small values of $\kappa$ and is converted to
a second order phase transition when $\kappa$ is increased beyond a 
finite tricritical value. 

\section{Conclusions}
In this paper we have, in addition to a number of bench-mark models, considered 
various lattice gauge models in $2+1$ Euclidean dimensions, with particular emphasis 
on the compact abelian Higgs model with integer gauge charge $q \geq 2$. We have 
seen that that for all $q \geq 2; q \neq 3$, the phase transition for the model 
in Eq. (\ref{Model1}) is second order such that the entire phase
transition line 
$\beta_{\rm{PT}}(\kappa)$ 
is critical, i.e. $\beta_{\rm{PT}}(\kappa)=\beta_{c}(\kappa)$. The exponents 
$\alpha$ and $\nu$ which are given by the $3DXY$ values in the limit 
$\kappa \gg \beta$, and global $Z_q$ values in the limit $\beta \gg \kappa$. 
For the special case $q=2$ and for intermediate values of $\beta$ and $\kappa$ 
on the critical line, the exponents vary continuously 
from the value $(1+\alpha)/\nu=1.76$ in the $Z_2$-limit $\beta \gg \kappa$ to
the value $(1+\alpha)/\nu=1.467$ in the $U(1)$ limit $\kappa \gg \beta$. This
constitutes a rare example of  a {\it fixed-line} theory  with non-universal
exponents  in a three dimensional system \cite{Note3}. What the connection
to the work of Refs. \onlinecite{KNS,KNS1} is, where a Kosterlitz-Thouless like phase 
transition of unbinding of monopoles in a three dimensional compact $U(1)$ 
gauge theory was found, remains to be investigated. More work is also required 
to elucidate the special role of $q=3$ in the compact abelian Higgs lattice 
gauge model. Note that this is fundamentally different from what happens in 
the $q$-state Potts model, to which the $Z_q$ spin model is equivalent only for 
$q=2,3$. In constrast to the compact abelian Higgs model, the $q$-state Potts 
model has a first order phase transition for {\it all } $q > 2.625$ 
\cite{Gliozzi}. 

Based on the connection between the model Eq. (\ref{Model1}) for $q=2$ and
a recently proposed model of fractionalized phases in strongly correlated
systems \cite{Motrunic} which is essentially given by Eq. (\ref{Model1}), 
we propose that the universality class of the putative quantum phase transition
from a Mott-Hubbard insulator to a charge-fractionalized insulator which the
model is supposed to describe, is in the universality class of the $q=2$ compact 
abelian Higgs model characterized by a fixed line of nonuniversal critical 
exponents varying continuously between the values $(1 + \alpha)/\nu = 1.76$ in 
the limit $\beta \to \infty$, $\kappa$ finite, and $(1 + \alpha)/\nu = 1.467$ 
in the limit $\kappa \to \infty$, $\beta$  finite. However, we have found that
over a significant portion of the critical line of the $q=2$ compact
abelian Higgs model, the exponents take on values consistent with those of 
the $3D$ Ising model. If microscopic models describing such Mott insulator-
fractionalized insulator transitions yield the $q=2$ compact abelian Higgs model
with sufficiently large values of $\beta$ and small values of $\kappa$ as an 
effective theory, then the resulting insulator-insulator transition is in the
$3D$ Ising universality class \cite{Motrunic}.  

%Another interesting point for further study, will be to investigate whether a 
%phase transition may occur within the phase below the line $\beta_c(\kappa)$. 
%Preliminary numerical studies on the case $q=2$ shows a pronounced peak in 
%the second moment of the action as $\kappa$ is varied in this region. 
%It is known that for $d=3$ no Coulomb phase exists. Consistent with this,
%we have found no sign of a first order phase transition, or indeed any ordinary 
%second order phase transition, for the system sizes we have considered so far 
%in this part of the parameter space. This may however be an artifact of too 
%small system sizes. It is also possible that the situation may be analogous 
%to what is known to occur in the Kosterlitz-Thouless transition in the $2DXY$ 
%model \cite{KT}, where the second moment has a broad peak well above the 
%critical temperature, and where the non-analyticity  is an essential singularity 
%completely indescernible in the second moment. This may  provide a connection to 
%the results obtained in Ref. \onlinecite{KNS}, where a Kosterlitz-Thouless like 
%phase transition of unbinding of monopoles in three dimensions in a compact 
%three dimensional abelian Higgs model was found. 
It 
%thus 
appears that the study of matter-coupled compact gauge theories would benefit 
greatly from a precise formulation of a non-local order parameter
represented by generalized rigidity for such models, as a 
substitute for the Wilson- or Polyakov loops which have proved useful 
in the pure gauge theories, but which are always perimeter-law bounded
in the matter coupled case when the symmetry group of the matter field is
a subgroup of the symmetry group for the gauge field.

\acknowledgments

A. S. and F.S.N. acknowledge support from the Norwegian Research Council 
(NFR), Grant No. 148825/432 and from DFG Sonderforschungsbereich 290, 
respectively. J. S and E. S. acknowledge NTNU for support through university 
fellowships. A. S. also thanks Prof. H. Kleinert and the Institut f{\"u}r 
Theoretische Physik,  Freie Universit{\"a}t Berlin for hospitality. We thank 
C. Mudry  and M. Hasenbusch for useful communications. NFR and the NTNU are 
acknowledged for support through computing time at the Norwegian High Performance 
Computing Centre (NOTUR). All computations were performed at the Norwegian 
High Performance Computing Centre on an Origin SGI 3800.  We particularly 
thank Dr. K. Rummukainen for providing the software with which we have 
analyzed our simulation data, and for many useful and enlightening 
discussions.  We also thank Nordita for hospitality.


\begin{thebibliography}{100}
\bibitem{Note_S} Note that 
this contrasts with the results of \cite{Sedgewick}, where the $U(1)$ symmetry group of 
the matter field is not contained in the $Z_2$ symmetry group of the gauge-field. Hence, 
the results of \cite{Nussinov} are not relevant. In \cite{Sedgewick}, the Polyakov loop 
could therefore be  used with success to map out the phase-structure of the theory, even 
if the gauge theory considered in  \cite{Sedgewick} is matter-coupled.
\end{thebibliography}
\end{document}